\documentclass[11pt]{article}
\pdfoutput=1

\usepackage{jheppub}
\usepackage[T1]{fontenc}

\usepackage{amssymb}
\usepackage{amsfonts}
\usepackage{amsbsy}
\usepackage{amsmath}
\usepackage{amsthm}
\usepackage{graphicx}
\usepackage{epstopdf}
\newcommand{\new}[1]{{\em #1}}

\def\Z{\mathbb{Z}}
\def\F{\mathbb{F}}

\def\C{\mathbb{C}}

\def\P{\mathbb{P}}

\def\n3a{t}

\def\ge{{\mathfrak{e}}}
\def\gso{{\mathfrak{so}}}
\def\gsu{{\mathfrak{su}}}

\def\gf{{\mathfrak{f}}}
\def\gg{{\mathfrak{g}}}

\def\ho{h^{1, 1}}
\def\htt{h^{2, 1}}
\newcommand{\eq}[1]{(\ref{#1})}

\title{6D F-theory models
and elliptically fibered Calabi-Yau threefolds over semi-toric base surfaces}

\author{
Gabriella Martini and Washington Taylor\\
Center for Theoretical Physics\\
Department of Physics\\
Massachusetts Institute of Technology\\
77 Massachusetts Avenue\\
Cambridge, MA 02139, USA\\[0.7cm]
{\tt gmartini} {\rm at} {\tt mit.edu},
{\tt wati} {\rm at} {\tt mit.edu}
}

\preprint{MIT-CTP-4448}

\abstract{We carry out a systematic study of a class of 6D F-theory
  models and associated Calabi-Yau threefolds that are constructed
  using base surfaces with a generalization of toric structure.  In
  particular, we determine all smooth surfaces with a structure
  invariant under a single $\C^*$ action (sometimes called
  ``T-varieties'' in the mathematical literature) that can act as
  bases for an elliptic fibration with section of a Calabi-Yau
  threefold.  We identify 162,404 distinct bases, which include as a
  subset the previously studied set of strictly toric bases.
  Calabi-Yau threefolds constructed in this fashion include examples
  with previously unknown Hodge numbers.  There are also bases over
  which the generic elliptic fibration has a Mordell-Weil group of
  sections with nonzero rank, corresponding to non-Higgsable $U(1)$
  factors in the 6D supergravity model; this type of structure does
  not arise for generic elliptic fibrations in the purely toric
  context.}

\begin{document}
\maketitle

\flushbottom

%--------------------------------
\section{Introduction}

Since the early days of string theory, much effort has been devoted to
understanding the geometry of string compactifications.  Calabi-Yau
threefolds are one of the most central and best-studied classes of
compactification geometries.  These manifolds can be used to
compactify ten-dimensional superstring theories to give
four-dimensional supersymmetric theories of gravity and gauge fields
\cite{GSW, Polchinski}.  Calabi-Yau
 threefolds
that admit an elliptic fibration with section can also be used to compactify F-theory to give
six-dimensional supergravity theories \cite{Vafa-F-theory,
  Morrison-Vafa}.

While mathematicians and physicists have used many methods to
construct and study Calabi-Yau threefolds (see \cite{Davies} for a
recent review), one of the main approaches that has been fruitful for
systematically classifying large numbers of Calabi-Yau geometries is
the mathematical framework of \emph{toric geometry} \cite{Fulton}.
The power of toric geometry is that many geometric features of a space
are captured in a simple combinatorial framework that lends itself to
straightforward calculations for many quantities of interest.  An
approach was developed by Batyrev \cite{Batyrev} for describing
Calabi-Yau manifolds as hypersurfaces in toric varieties in terms of
the combinatorics of reflexive polytopes.  Kreuzer and Skarke have
identified some 473.8 million four-dimensional reflexive polytopes
that can be used to construct Calabi-Yau threefolds in this way
\cite{Kreuzer-Skarke}.

In this paper, following \cite{clusters}, we study a class of
spaces that is more general than the set of toric varieties, but
retains some of the combinatorial simplicity of toric geometry.  This
allows us to construct a large class of elliptically fibered
Calabi-Yau threefolds that need not have any realization in a strictly
toric language.  In particular, we focus on complex surfaces that
admit at least one $\C^*$ action but not necessarily the action of a
product $\C^*\times \C^*$ needed for a full toric structure.  Complex
algebraic varieties of dimension $n$ that admit an effective action of
$(\C^*)^k$ are known in the mathematical literature as
``T-varieties''.  In this paper we simply use the term ``$\C^*$-surface''
to denote a surface with a $\C^*$ action, in part to avoid confusion
with the plethora of ``T-'' objects already filling the string theory
literature (``T-duality'', ``T-folds'', ``T-branes'', etc.)  A review
of some of the mathematical results and literature on more general
T-varieties can be found in \cite{T-varieties}.

The primary focus of this paper is the systematic classification and
study of all smooth $\C^*$-surfaces that can act as bases for an
elliptic fibration with section where the total space is a Calabi-Yau
threefold.  We use these surfaces to compactify F-theory to six
dimensions.  The close correspondence between the geometry of the
F-theory compactification surface and the physics of the associated 6D
supergravity theory provides a powerful tool for understanding both
geometry and physics.  A general characterization of smooth base
surfaces\footnote{Throughout this paper we use the term ``base
  surface'' as shorthand for ``surface that can act as the base of an
  elliptically-fibered Calabi-Yau threefold with section''.} (not
necessarily toric or $\C^*$-) was given in \cite{clusters}.  The basic
idea is that any base can be classified by the intersection structure
of effective irreducible divisors of self-intersection $-2$ or below.
The intersection structure of the base directly corresponds to the
generic nonabelian gauge group in the ``maximally Higgsed'' 6D
supergravity theory from an F-theory construction.  There are strong
constraints on the intersection structures that can arise; these
constraints made possible an explicit enumeration of all toric base
surfaces in \cite{toric}, and the geometry of the associated
Calabi-Yau threefolds was explored in \cite{WT-Hodge}.  In this paper
we construct and enumerate the more general class of $\C^*$-bases and
explore the corresponding Calabi-Yau geometries.  In particular, we
identify some (apparently) new Calabi-Yau threefolds including some
threefolds with novel properties.

In Section \ref{sec:bases} we describe the class of $\C^*$-base
surfaces.  We summarize the results of the complete enumeration of
these bases in Section \ref{sec:enumeration}.  The corresponding
Calabi-Yau threefolds, including some models with interesting new
features, are described in Section \ref{sec:features}.  Section
\ref{sec:conclusions} contains concluding remarks.
Appendices~\ref{sec:rules}--\ref{sec:appendix-abelian} contain tables
of useful data.

\section{$\C^*$-surfaces and F-theory vacua}
\label{sec:bases}

We begin by briefly summarizing the results of \cite{clusters,
  toric}.  The methods used here are closely related to those
developed in these papers.

\subsection{6D F-theory models}

A 6D F-theory model is defined by a Calabi-Yau threefold that is an
elliptic fibration (with section) over a complex base surface $B$.
The structure of the resulting 6D supergravity theory is determined by
the geometry of $B$ and the elliptic fibration \cite{Morrison-Vafa}.
In particular, the number of tensor multiplets $T$ in the 6D theory is
related to the topology of the base $B$ through $T = h^{1, 1} (B) -1$.
The elliptic fibration can be described by a Weierstrass
model over $B$
\begin{equation}
y^2 = x^3 + fx + g \,,
\label{eq:Weierstrass}
\end{equation}
where $f, g$ are sections of the line bundles ${\cal O} (-4K), 
{\cal O}(-6K),$ with $K$ the
canonical class of $B$.  Codimension one vanishing loci of $f,
g$, and the discriminant locus $\Delta = 4f^3+27g^2$, where the
elliptic fibration becomes singular, give rise to vector multiplets
for a nonabelian gauge group in the 6D theory.  Codimension two
vanishing loci give matter in the 6D theory; the matter lives in a
representation of the nonabelian gauge group that is determined by the
geometry.
For more detailed background regarding the relation of 6D physics to F-theory geometry,
see the reviews  \cite{Morrison-TASI, Denef-F-theory, WT-TASI}; a
recent systematic analysis of these 6D models
from the M-theory point of view appears in
\cite{b-Grimm}).

\subsection{General classification of 6D F-theory
base surfaces}

\begin{figure}
\begin{center}
\begin{picture}(200,130)(- 93,- 55)
%\grid
\thicklines
\put(-175, 25){\line(1,0){50}}
%\put(-150,32){\makebox(0,0){$-m\leq -3$}}
\put(-150, 47){\makebox(0,0){\small $-m \in$}}
\put(-150, 32){\makebox(0,0){\small  $\{-3, -4, \ldots, -8, -12\}$}}
%\put(-150, 18){\makebox(0,0){$(m = 3, 4, 5, 6, 7, 8, 12)$}}
\put(-150,-33){\makebox(0,0){\small $\gsu(3), \gso(8), \gf_4$}}
\put(-150,-47){\makebox(0,0){\small $\ge_6, \ge_7, \ge_8$}}
\put(-70,55){\line(1,-1){40}}
\put(-30,35){\line(-1,-1){40}}
\put(-50,45){\makebox(0,0){-3}}
\put(-50,5){\makebox(0,0){-2}}
\put(-50,-40){\makebox(0,0){\small $\gg_2 \oplus \gsu(2)$}}
\put(30,70){\line(1,-1){40}}
\put(30,20){\line(1,-1){40}}
\put(70,45){\line(-1,-1){40}}
\put(45,65){\makebox(0,0){-3}}
\put(44,31){\makebox(0,0){-2}}
\put(60, 0){\makebox(0,0){-2}}
\put(50,-40){\makebox(0,0){\small $\gg_2 \oplus \gsu(2)$}}
\put(130,70){\line(1,-1){40}}
\put(130,20){\line(1,-1){40}}
\put(170,45){\line(-1,-1){40}}
\put(145,65){\makebox(0,0){-2}}
\put(144,31){\makebox(0,0){-3}}
\put(160, 0){\makebox(0,0){-2}}
\put(150,-40){\makebox(0,0){\small $\gsu(2) \oplus \gso(7) \oplus \gsu(2)$}}
\end{picture}
\end{center}
\caption[x]{\footnotesize    Clusters of intersecting
  curves that must carry a nonabelian gauge group factor.  For each
cluster the corresponding gauge algebra is noted and the gauge algebra and
number of charged matter hypermultiplet are listed in Table~\ref{t:clusters}}
\label{f:clusters}
\end{figure}
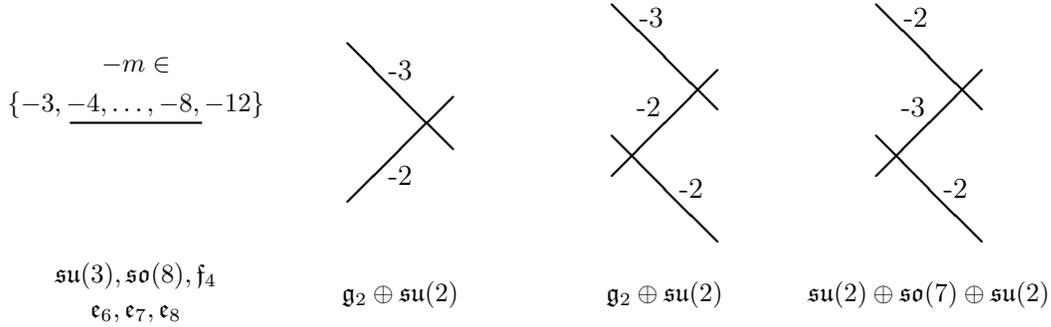

A general approach to systematically classifying base surfaces $B$ for
6D F-theory compactifications was developed in \cite{clusters}.
This approach is based on identifying irreducible components in the
structure of effective divisors on $B$, composed of
intersecting combinations of curves of negative self-intersection over
which the generic elliptic fibration is singular.
Each such irreducible component corresponds to a ``non-Higgsable cluster'' of
gauge algebra summands and (in some cases) charged matter that appears
in the 6D supergravity model arising from an F-theory compactification
on a generic elliptic fibration over $B$; the term ``non-Higgsable
cluster'' (which we sometimes abbreviate as ``cluster'') refers to the
fact that for any such configuration there are no matter fields that
can lead to a Higgsing of the corresponding nonabelian gauge group.
For example, a single irreducible effective curve in $B$ with
self-intersection $-4$ corresponds to an $\gso(8)$ term in the gauge
algebra with no charged matter.  The simplest example of this occurs
for F-theory on the Hirzebruch surface $\F_4$
\cite{Morrison-Vafa}.  There are strong geometric
constraints (discussed further below) on the configurations of curves
that can live in a base supporting an elliptic fibration.  The set of
possible non-Higgsable clusters
is thus rather
small.  A complete list is given in  Table~\ref{t:clusters}, and
depicted in Figure~\ref{f:clusters}.  These
clusters are described by configurations of curves having
self-intersection $-2$ or less, with one curve having
self-intersection $-3$ or below.  
General configurations of $-2$
curves with no curves of self-intersection $-3$ or below
are also possible.  These clusters carry no gauge group, and generally
represent limiting points of bases without these $-2$ curves; for example,
the Hirzebruch surface $\F_2$ contains a single isolated
$-2$ curve, and is a
limit of the surface $\F_0$.

\begin{table}
\begin{center}
\begin{tabular}{| c | 
c |c |
}
\hline
Cluster & gauge algebra &  $H_{\rm charged}$ 
\\
\hline
(-12) &$\ge_8$ & 0 \\
(-8) &$\ge_7$&  0 \\
(-7) &$\ge_7$& 28 \\
(-6) &$\ge_6$&   0 \\
(-5) &$\gf_4$&   0 \\
(-4) &$\gso(8) $&  0 \\
(-3, -2, -2)  &  $\gg_2 \oplus \gsu(2)$&  8\\
(-3, -2) &  $\gg_2 \oplus \gsu(2)$ &8 \\
(-3)& $\gsu(3)$ &  0 \\
(-2, -3, -2) &$\gsu(2) \oplus \gso(7) \oplus
\gsu(2)$&16 \\
(-2, -2, \ldots, -2) & no gauge group & 0 \\
\hline
\end{tabular}
\end{center}
\caption[x]{\footnotesize Allowed ``non-Higgsable clusters'' of
  irreducible effective divisors with self-intersection $-2$ or below,
  and corresponding contributions to the gauge algebra and matter
  content of the 6D theory associated with F-theory compactifications
  on a generic elliptic fibration (with section) over a base
  containing each cluster.}
\label{t:clusters}
\end{table}

The restriction on the types of allowed clusters comes from the
constraint that $f, g$ cannot vanish to degrees $4, 6$ respectively on any curve or
at any point in $B$.  If the degree of vanishing is too high on a
curve, there is no way to construct a Calabi-Yau from the elliptic
fibration.  If the vanishing is too high at a point, the point must be
blown up to form another base $B'$ that supports a Calabi-Yau.  The
vanishing degrees of $f, g$ on a curve can be determined from the
\emph{Zariski decomposition} of $-nK$: using the fact that any effective divisor
$A$ that has a negative intersection with an effective irreducible $B$
having negative self-intersection must contain $B$ as a component ($A
\cdot B < 0, B \cdot B < 0 \Rightarrow A = B + C$ with $C$ effective),
the degree of vanishing of $ A = -4K, -6K, -12K$ on any irreducible
effective curve or combination of curves can be computed; the degree
of vanishing at a point where two curves intersect is simply the sum
of the degrees of vanishing on the curves.  The details of this
computation for general combinations of intersecting divisors are
worked out in \cite{clusters}. 

All effective irreducible curves of self-intersection $-2$ or below in
the base appear in the clusters described above.  Furthermore, again
because of restrictions on the degrees of $f, g$ on curves and intersection
points, only certain combinations of the allowed
clusters can be connected by $-1$ curves in $B$.  For example, 
if a $-1$ curve intersects
a $-12$
curve, then the $-1$ curve cannot intersect any other curve contained
in a non-Higgsable
cluster except a $-2$ curve contained in a $(-2, -2, -3)$
cluster. 
A complete table of the possible combinations
of clusters that can be connected by a $-1$ curve is given in
\cite{clusters}.  The part of that table that is relevant for this
paper is reproduced in Appendix~\ref{sec:rules}.

For any base surface $B$, the structure of non-Higgsable clusters
determines the minimal nonabelian gauge group and matter content of
the 6D supergravity theory corresponding to a generic elliptic
fibration with section over $B$.  For each distinct base $B$ there can
be a wide range of models with different nonabelian gauge groups,
which can be realized by tuning the Weierstrass model to increase the
degrees of $f, g$ over various divisors.  For example, for the
simplest base surface $B = \P^2$, there are thousands of branches of
the theory with different nonabelian gauge group and matter content,
some of which are explored in \cite{0, Braun-0}.  Each of these
branches corresponds to a different Calabi-Yau threefold after the
singularities associated with the nonabelian gauge group factors are
resolved.  But for each of these models, by maximally Higgsing matter
fields in the supergravity theory the gauge group can be completely
broken and the geometry becomes that of a generic elliptic fibration
over $\P^2$.  Focusing on the base surface $B$ dramatically simplifies
the problem of classifying F-theory compactifications and elliptically
fibered Calabi-Yau threefolds, by removing the additional complexity
associated with the details of specific Weierstrass models and
associated fibrations.

The mathematics of minimal surface theory \cite{bhpv, Reid-chapters}
gives a simple picture of how the set of allowed base surfaces $B$ are
connected, unifying the space of 6D supergravity theories that arise
from F-theory.  All smooth bases\footnote{aside from the Enriques
surface, which gives rise to a simple 6D model with no gauge group or
matter and is connected in a more complicated way to the branches
associated with other bases.}  $B$ for 6D F-theory models can be
constructed by blowing up a finite set of points on one of the minimal
bases $\F_m\ (0 \leq m \leq 12, m \neq 1)$ or $\P^2$ \cite{Grassi}.
The number of distinct topological types for $B$ is finite
\cite{Gross, KMT-II}.  In principle, all smooth bases $B$ can be
systematically constructed by successively blowing up points on the
minimal bases allowing only the clusters of irreducible divisors from
Table~\ref{t:clusters}.  In \cite{toric}, the complete set of toric
bases was constructed in this fashion, and in this paper we carry out
the analogous construction for $\C^*$-bases.

\subsection{Toric base surfaces}
\label{sec:toric}

\begin{figure}
\setlength{\unitlength}{.8pt}
\begin{center}
\begin{picture}(200,130)(-93, -25)
% First sub fig
\put(-235, 60){\line(1,0){80}}
\put(-235, 0){\line(1,0){80}}
\put(-225, 70){\line( 0, -1){80}}
\put( -225,60){\makebox(0,0){\includegraphics[width=0.4cm]{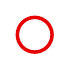}}}
\put(-165, 70){\line( 0, -1){80}}
\put(-195, 50){\makebox(0,0){$+2$}}
\put(-195, 72){\makebox(0,0){\small{$C_{1}(D_{0})$}}}
\put(-195, -11){\makebox(0,0){\small{$C_{3}(D_{\infty})$}}}
\put(-195, 10){\makebox(0,0){$-2$}}
\put(-217,30){\makebox(0,0){$0$}}
\put(-235,30){\makebox(0,0){\small{$C_{4}$}}}
\put(-176,30){\makebox(0,0){$0$}}
\put(-154,30){\makebox(0,0){\small $C_{2}$}}
%\put(-140, 70){\makebox(0,0){{\large $D_{0}$}}}
%\put(-140, -10){\makebox(0,0){{\large $D_{\infty}$}}}
\put(-135, 30){\vector( 1, 0){25}}
% Second subfig
\put(-80, 60){\line(1,0){75}}
\put(-80, 0){\line(1,0){75}}
\put(-67, 68){\line( -1, -2){23}}
\put( -71,60){\makebox(0,0){\includegraphics[width=0.4cm]{circle.pdf}}}
\put(-67, -8){\line( -1, 2){23}}
\put(-15, 70){\line( 0, -1){80}}
\put(-40, 70){\makebox(0,0){$+1$}}% was 2
\put(-40, -10){\makebox(0,0){$-2$}}
\put(-8,30){\makebox(0,0){$0$}}
\put(-93,50){\makebox(0,0){$-1$}}
\put(-93,10){\makebox(0,0){$-1$}}
\put(7, 30){\vector( 1, 0){25}}
% Third Sub fig
\put(70, 60){\line(1,0){66}}
\put(70, 0){\line(1,0){66}}
\put(81, 66){\line( -4, -5){22}}
\put(81, -6){\line( -4, 5){22}}
\put(122,0){\makebox(0,0){\includegraphics[width=0.4cm]{circle.pdf}}}
\put(62, 48){\line( 0, -1){36}}
\put(122, 70){\line( 0, -1){80}}
\put(92, 70){\makebox(0,0){$0$}}% was 2, changed below also.
\put(92, -10){\makebox(0,0){$-2$}}
\put(129,30){\makebox(0,0){$0$}}
\put(59,55){\makebox(0,0){$-1$}}
\put(59,4){\makebox(0,0){$-1$}}
\put(49,30){\makebox(0,0){$-2$}}
\put(138, 30){\vector( 1, 0){25}}
% Fourth Sub Fig
\put(200, 60){\line(1,0){62}}
\put(200, 0){\line(1,0){62}}
\put(211, 65){\line( -4, -5){22}}
\put(211, -5){\line( -4, 5){22}}
\put(192, 48){\line( 0, -1){36}}
\put(248, 65){\line( 1, -2){21}}
\put(248, -5){\line( 1, 2){21}}
\put(226, 72){\makebox(0,0){$D_{0}$}}
\put(226, 50){\makebox(0,0){$0$}}
\put(225, 10){\makebox(0,0){$-3$}}
\put(227, -10){\makebox(0,0){$D_{\infty}$}}
\put(189,55){\makebox(0,0){$-1$}}
\put(189,4){\makebox(0,0){$-1$}}
\put(179,30){\makebox(0,0){$-2$}}
\put(268,50){\makebox(0,0){$-1$}}
\put(268,10){\makebox(0,0){$-1$}}
\end{picture}
\end{center}
\caption[x]{\footnotesize  Toric base surfaces for 6D F-theory models
  produced by blowing up a sequence of points on $\F_2$.}
\label{f:loop}
\end{figure}

The set of toric base surfaces form a subset of the more general class
of $\C^*$-bases that we focus on in this paper.  
The structure of toric bases is relatively simple and
provides a useful foundation for
our analysis of $\C^*$-bases.

A toric surface can be described alternatively in terms of the
\new{toric fan} or in terms of the set of toric effective divisors.
The fan for a compact toric surface is defined \cite{Fulton} by a sequence of
vectors $v_1, \ldots, v_k \in N =\Z^2$ defining 1D cones, or {\it
rays}, along with 2D cones spanned by vectors $v_i, v_{i +1}$
(including a 2D cone spanned by $v_k, v_1$; all related conditions
include this periodicity though we do not repeat this explicitly in
each case), and the 0D cone at the origin.  The origin represents the
torus $(\C^*)^2$, while the 1D rays represent divisors and the 2D
cones represent points in the compact toric surface.  The surface is
smooth if the rays $v_i, v_{i +1}$ defining each 2D cone span a unit
cell in the lattice.  The irreducible effective toric divisors are a
set of curves\footnote{Note that the notation and ordering
used here for curves and associated rays in a toric base differs from
that used in \cite{toric}.} $C_i, i = 1, \ldots, k$, associated with the vectors
$v_i$.  The divisors have self-intersection $C_i \cdot C_i = n_i$,
where $-nv_i =v_{i-1} + v_{i +1}$, and nonvanishing pairwise
intersections $C_i \cdot C_{i +1} = C_k \cdot C_1 = 1$.
These
divisors can be depicted graphically as a loop (See
Figure~\ref{f:loop}).  The sets of consecutive divisors of
self-intersection $n_i \leq -2$ in the loop are constrained by the
cluster analysis of \cite{clusters} to contain only the sequences
(with either orientation) $(-3, -2), (-3, -2, -2), (-2, -3, -2)$,
$(-m)$, with $3 \leq m \leq 12$, and $(-2, \ldots, -2)$ with any
number of $-2$ curves.

All smooth toric base surfaces (aside from $\P^2$) can be constructed
by starting with a Hirzebruch surface $\F_m$ that is associated with
the divisor self-intersection sequence $[n_1, n_2, n_3, n_4] = [m, 0,
-m, 0]$ ($v_1 = (0,1), v_2 = (1, 0), v_3 = (0, -1), v_4 = (-1, -m)$),
and blowing up a sequence of intersection points between adjacent
divisors.  Blowing up the intersection point between divisors $C_i$
and $C_{i +1}$ gives a new toric base with a -1 curve inserted between
these divisors, and self-intersections of the previously intersecting
divisors each reduced by one (See Figure~\ref{f:loop}).  Such blow-ups
are the only ones possible that maintain the toric structure by
preserving the action of $(\C^*)^2$ on the base.  Each Hirzebruch
surface can be viewed as a $\P^1$ bundle over $\P^1$.  The divisors of
self-intersection $\pm m$ can be viewed as sections $\Sigma_\pm$ of
this bundle, which in a local coordinate chart $(z, w) \in\C \times\C$
are at the points $(z,0), (z, \infty)$ in the fibers, while the
divisors of self-intersection $0$ can be viewed as fibers $(0, w)$ and
$(\infty, w)$.  All points that can be blown up while maintaining the toric structure are located at the points invariant under the $(\C^*)^2$ action: $(0,
0), (0, \infty), (\infty, 0), (\infty, \infty)$ (or in exceptional
divisors produced when these points are blown up).
In particular, any smooth toric base surface  that supports an
elliptically fibered Calabi-Yau threefold
has a description in terms of a
closed loop of divisors containing divisors $D_0, D_\infty$ associated
with the divisors
$C_1, C_3$ in the original Hirzebruch surface (these
are the sections $\Sigma_\pm$) and two linear chains of divisors connecting
$D_0$ to $D_\infty$.  These chains are formed from the 
divisors $C_2$ and $C_4$ in the original Hirzebruch
surface, along with all exceptional divisors from blowing up points on
these original curves.  This is illustrated in Figure~\ref{f:loop}.

Note that some smooth toric base surfaces can be formed in different
ways from blow-ups of Hirzebruch surfaces, so that different divisors
$D_0, D_\infty$
play the role of the sections $\Sigma_\pm$.  In enumerating all toric
base surfaces, such duplicates must be eliminated by considering
equivalences of the loops of self-intersection numbers up to rotation
and reflection.

\subsection{$\C^*$-base surfaces}
\label{sec:semi-toric}

A more general class of bases $B$ was described in \cite{toric}.
Generalizing the toric construction by allowing blow-ups at arbitrary
points $(z, 0), (z, \infty)$ can give rise to an arbitrary number of
fibers containing curves of negative self-intersection, while
maintaining the $\C^*$ action ($1 \times\C^*$) that leaves the
sections $\Sigma_\pm$ invariant.  The resulting structure is closely
analogous to that of the toric bases described above, except that
there can be more than two chains of intersecting divisors connecting
$D_0, D_\infty$, associated with distinct blown-up fibers of the original
Hirzebruch surface.  Graphically, the intersection structure of
effective irreducible divisors for such a base can be depicted as the
pair of horizontal divisors $D_0, D_\infty$ (associated with
$\Sigma_\pm$ in the original Hirzebruch surface), along with an arbitrary number $N$ of chains of divisors
$D_{i, j}, i = 1, \ldots, N$ connecting $D_0, D_\infty$.  The $i$th
chain is realized by blowing up a fiber in $\F_m$, and contains
divisors $D_{i, j}, j = 1, \ldots, k_i$, with self-intersections and
adjacent intersections as described above ($D_{i, j} \cdot D_{i, j} =
n_{i, j}  < 0$, $D_{i, j} \cdot D_{i, j + 1} = 1$) except that
$D_{i,1} \cdot D_0 = 1, D_{i, k_i} \cdot D_\infty = 1,$ and $D_{i,1}
\cdot D_{i, k_i} = 0$ unless $k_i = 2$.  (See
Figure~\ref{f:semi-toric}.)  We refer to bases of this form as
\new{$\C^*$-bases}.  In this language, the toric bases correspond to
those $\C^*$-bases with $N = 2$ (or fewer) chains.  The divisor
intersection structure of a $\C^*$-base is constrained by the set of
allowed clusters found in \cite{clusters} in a similar way to toric
bases, but with additional possibilities associated with the existence
of multiple connections to the divisors $D_0, D_\infty$ that provide
branchings in the intersection structure.  For example, if $D_0$ has
self-intersection $D_0 \cdot D_0 = -3$, and is connected to chains
with self-intersections $(n_{1, 1}, n_{1, 2}, \ldots) = (-2, -1,
\ldots)$, $(n_{2, 1}, n_{2, 2}, \ldots) = (-2, -1, \ldots)$, then any
further chains $i > 2$ must satisfy $n_{i,1}= -1, n_{i, 2} \geq -4$.
The complete set of constraints on how divisors can be connected is described in
Appendix~\ref{sec:rules}, including some additional constraints
beyond those described in \cite{clusters}
related to branchings on curves of self-intersection $-n < -1$.

Note that the value of $T = h^{1, 1} (B) -1$ can be determined
directly from the intersection structure of a $\C^*$-base.  Each
blow-up adds one to $T$.  Each blow-up along a new fiber increases $N$
by one and creates a new $(-1, -1)$ chain with $k_N = 2$, while each
blow-up on an intersection  in chain $i$ increases
$k_i$ by 1.  Matching with the Hirzebruch surfaces, which have $N = 0,
T = 1$, we have
\begin{equation}
T = h^{1, 1} (B) -1 =(\sum_{i = 1}^{N}  k_i) -N + 1 \,.
\label{eq:t-equation}
\end{equation}
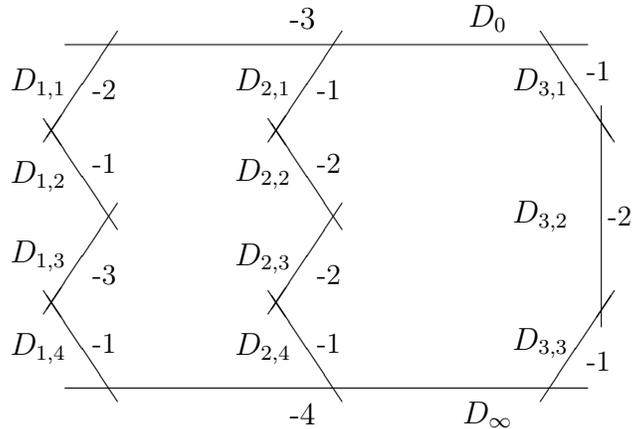
\begin{figure}
\begin{center}
\begin{picture}(200,160)(- 100,- 80)
\put(-100, 65){\line(1, 0){198}}
\put(-100, -65){\line(1, 0){198}}
\put(60,75){\makebox(0,0){{\large$D_{0}$}}}% added
%\put(90,-75){\makebox(0,0){{\large$D_{\infty}$}}}
\put(60,-75){\makebox(0,0){{\large$D_{\infty}$}}}
\put(-10,75){\makebox(0,0){{\large-3}}}% was 2
\put(-10,-75){\makebox(0,0){{\large-4}}}% was 3
% Fiber 1
\put(-80, 70){\line( -2, -3){28}}
\put(-80, -5){\line( -2, 3){28}}
\put(-80, 5){\line( -2, -3){28}}
\put(-80, -70){\line(-2, 3){28}}
\put(-110,50){\makebox(0,0){{\large$D_{1,1}$}}}
\put(-110,15){\makebox(0,0){{\large$D_{1,2}$}}}
\put(-110,-15){\makebox(0,0){{\large$D_{1,3}$}}}
\put(-110,-50){\makebox(0,0){{\large$D_{1,4}$}}}
\put(-85,48){\makebox(0,0){-2}}
\put(-85,20){\makebox(0,0){-1}}
\put(-85,-22){\makebox(0,0){-3}}
\put(-85,-48){\makebox(0,0){-1}}

% Fiber 2
\put(5, 70){\line( -2, -3){28}}
\put(5, -5){\line( -2, 3){28}}
\put(5, 5){\line( -2, -3){28}}
\put(5, -70){\line(-2, 3){28}}
\put(-25,50){\makebox(0,0){{\large$D_{2,1}$}}}
\put(-25,16){\makebox(0,0){{\large$D_{2,2}$}}}
\put(-25,-16){\makebox(0,0){{\large$D_{2,3}$}}}
\put(-25,-50){\makebox(0,0){{\large$D_{2,4}$}}}
\put(0,48){\makebox(0,0){-1}}
\put(0,20){\makebox(0,0){-2}}
\put(0,-22){\makebox(0,0){-2}}
\put(-0,-48){\makebox(0,0){-1}}

% Fiber 3
\put(80, 70){\line( 2, -3){28}}
\put(80, -70){\line(2, 3){28}}
\put(103, 42){\line(0, -1){84}}
\put(80,50){\makebox(0,0){{\large$D_{3,1}$}}}
\put(80,0){\makebox(0,0){{\large$D_{3,2}$}}}
\put(80,-48){\makebox(0,0){{\large$D_{3,3}$}}}
\put(102,55){\makebox(0,0){-1}}
\put(110,0){\makebox(0,0){-2}}
\put(102,-55){\makebox(0,0){-1}}
\end{picture}
\end{center}
\caption[x]{\footnotesize  A $\C^*$-base $B$ is characterized by
irreducible effective  divisors $D_0, D_\infty$ connected by any
number of linear chains of divisors $D_{i, j}$ with intersections
obeying the cluster rules of \cite{clusters}.  All such bases can be
realized as multiple blow-ups of a Hirzebruch surface $\F_m$ that
preserve the action of a single $\C^*$ on $B$}
\label{f:semi-toric}
\end{figure}

\subsection{Curves of self-intersection -9, -10, and -11}

\begin{figure}
\setlength{\unitlength}{.83pt}
\begin{center}
\begin{picture}(180,160)(- 100,- 80)
\put(-30, 0){\vector(1,0){30}}
\put(-265, 30){\line(1, 0){225}}
\put(-265, -30){\line(1, 0){225}}
\put(-155,40){\makebox(0,0){0}}
\put(-155,-40){\makebox(0,0){-11}}
\put(-50,-40){\makebox(0,0){{\large$D_{\infty}$}}}
\put(-50,40){\makebox(0,0){{\large$D_{0}$}}}
% Fiber zero
\put(-255, 32){\line( -1, -3){12}}
\put(-255, -32){\line(-1, 3){12}}
\put(-268,15){\makebox(0,0){-1}}
\put(-268,-15){\makebox(0,0){-1}}

% Fiber one
\put(-235, 32){\line( -1, -3){12}}
\put(-235, -32){\line(-1, 3){12}}
\put(-248,15){\makebox(0,0){-1}}
\put(-248,-15){\makebox(0,0){-1}}

% Fiber two
\put(-215, 32){\line( -1, -3){12}}
\put(-215, -32){\line(-1, 3){12}}
\put(-228,15){\makebox(0,0){-1}}
\put(-228,-15){\makebox(0,0){-1}}
% Fiber three
\put(-195, 32){\line( -1, -3){12}}
\put(-195, -32){\line(-1, 3){12}}
\put(-208,15){\makebox(0,0){-1}}
\put(-208,-15){\makebox(0,0){-1}}
% Fiber four
\put(-175, 32){\line( -1, -3){12}}
\put(-175, -32){\line(-1, 3){12}}
\put(-188,15){\makebox(0,0){-1}}
\put(-188,-15){\makebox(0,0){-1}}
% Fiber five
\put(-155, 32){\line( -1, -3){12}}
\put(-155, -32){\line(-1, 3){12}}
\put(-168,15){\makebox(0,0){-1}}
\put(-168,-15){\makebox(0,0){-1}}
% Fiber six
\put(-135, 32){\line( -1, -3){12}}
\put(-135, -32){\line(-1, 3){12}}
\put(-148,15){\makebox(0,0){-1}}
\put(-148,-15){\makebox(0,0){-1}}
% Fiber seven
\put(-115, 32){\line( -1, -3){12}}
\put(-115, -32){\line(-1, 3){12}}
\put(-128,15){\makebox(0,0){-1}}
\put(-128,-15){\makebox(0,0){-1}}
% Fiber eight
\put(-95, 32){\line( -1, -3){12}}
\put(-95, -32){\line(-1, 3){12}}
\put(-108,15){\makebox(0,0){-1}}
\put(-108,-15){\makebox(0,0){-1}}
% Fiber nine
\put(-75, 32){\line( -1, -3){12}}
\put(-75, -32){\line(-1, 3){12}}
\put(-88,15){\makebox(0,0){-1}}
\put(-88,-15){\makebox(0,0){-1}}
% Fiber ten
\put(-55, 32){\line( -1, -3){12}}
\put(-55, -32){\line(-1, 3){12}}
\put(-68,15){\makebox(0,0){-1}}
\put(-68,-15){\makebox(0,0){-1}}
% Fiber eleven
\put(-45, 32){\line( 0, -1){64}}
\put(-50,0){\makebox(0,0){0}}
\put(-45,-30){\makebox(0,0){\includegraphics[width=0.4cm]{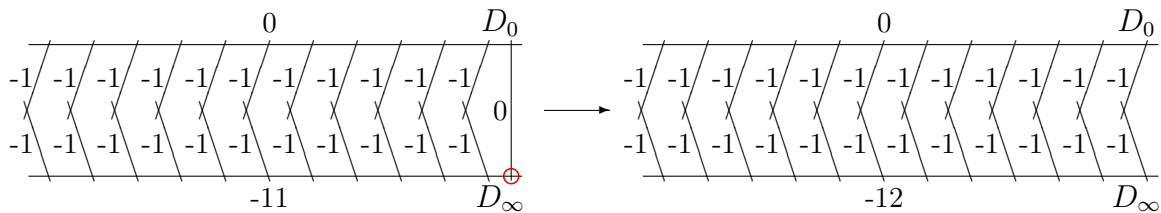}}}

% Second half
\put(15, 30){\line(1, 0){235}}
\put(15, -30){\line(1, 0){235}}
\put(125,40){\makebox(0,0){0}}
\put(125,-40){\makebox(0,0){-12}}
\put(240,-40){\makebox(0,0){{\large$D_{\infty}$}}}
\put(240,40){\makebox(0,0){{\large$D_{0}$}}}
% Fiber one
\put(25, 32){\line( -1, -3){12}}
\put(25, -32){\line(-1, 3){12}}
\put(12,15){\makebox(0,0){-1}}
\put(12,-15){\makebox(0,0){-1}}

% Fiber two
\put(45, 32){\line( -1, -3){12}}
\put(45, -32){\line(-1, 3){12}}
\put(32,15){\makebox(0,0){-1}}
\put(32,-15){\makebox(0,0){-1}}
% Fiber three
\put(65, 32){\line( -1, -3){12}}
\put(65, -32){\line(-1, 3){12}}
\put(52,15){\makebox(0,0){-1}}
\put(52,-15){\makebox(0,0){-1}}
% Fiber four
\put(85, 32){\line( -1, -3){12}}
\put(85, -32){\line(-1, 3){12}}
\put(72,15){\makebox(0,0){-1}}
\put(72,-15){\makebox(0,0){-1}}
% Fiber five
\put(105, 32){\line( -1, -3){12}}
\put(105, -32){\line(-1, 3){12}}
\put(92,15){\makebox(0,0){-1}}
\put(92,-15){\makebox(0,0){-1}}
% Fiber six
\put(125, 32){\line( -1, -3){12}}
\put(125, -32){\line(-1, 3){12}}
\put(112,15){\makebox(0,0){-1}}
\put(112,-15){\makebox(0,0){-1}}
% Fiber seven
\put(145, 32){\line( -1, -3){12}}
\put(145, -32){\line(-1, 3){12}}
\put(132,15){\makebox(0,0){-1}}
\put(132,-15){\makebox(0,0){-1}}
% Fiber eight
\put(165, 32){\line( -1, -3){12}}
\put(165, -32){\line(-1, 3){12}}
\put(152,15){\makebox(0,0){-1}}
\put(152,-15){\makebox(0,0){-1}}
% Fiber nine
\put(185, 32){\line( -1, -3){12}}
\put(185, -32){\line(-1, 3){12}}
\put(172,15){\makebox(0,0){-1}}
\put(172,-15){\makebox(0,0){-1}}
% Fiber ten
\put(205, 32){\line( -1, -3){12}}
\put(205, -32){\line(-1, 3){12}}
\put(192,15){\makebox(0,0){-1}}
\put(192,-15){\makebox(0,0){-1}}
% Fiber eleven
\put(225, 32){\line( -1, -3){12}}
\put(225, -32){\line(-1, 3){12}}
\put(212,15){\makebox(0,0){-1}}
\put(212,-15){\makebox(0,0){-1}}
% Fiber twelve
\put(245, 32){\line( -1, -3){12}}
\put(245, -32){\line(-1, 3){12}}
\put(232,15){\makebox(0,0){-1}}
\put(232,-15){\makebox(0,0){-1}}
\end{picture}
\end{center}
\caption[x]{\footnotesize The $\C^*$-base with $N = 11, n_0 =  0,
  n_\infty =  -11$ and 11 $(-1, -1)$ chains has a singular point on $D_\infty$
where the base must be blown up, giving the smooth base with $N = 12,
n_\infty = -12$, and 12 $(-1, -1)$ chains.}
\label{f:11-12}
\end{figure}

There is one further issue that arises in systematically classifying
toric and/or $\C^*$-bases.  As shown in \cite{clusters}, for
any base containing a curve $C$ of self-intersection $-9, -10,$ or
$-11$, there is a point on $C$ where the Weierstrass functions $f, g$
vanish to degrees $4, 6$, so that point in the base must be blown up
for the base to support an elliptically fibered Calabi-Yau threefold.  
If the curve $C$ is a divisor
associated with a section, $D_0$ or $D_\infty$, then blowing up this
point simply adds another $(-1, -1)$ chain to the $\C^*$-surface
(Figure~\ref{f:11-12}).
If $C$ is an element of one of the chains, however, then 
blowing up the point on $C$ generally
takes the base out of the class of
$\C^*$- or toric surfaces.

For strictly toric or $\C^*$-surfaces, therefore, we should not
include any bases containing curves of self-intersection -9, -10, or
-11.  For several reasons, however, we find it of interest to include
bases with such curves even when the blow-up leaves the context of $\C^*$-surfaces.  In particular, we are interested in exploring the widest
range of bases possible that can be systematically analyzed.  Thus,
while the bases arising from blowing up -9, -10, or -11 curves on
connecting chains generally takes us outside the $\C^*$ context we
have done a complete enumeration of surfaces including these
additional types of curves (which we refer to as ``not-strictly
$\C^*$-bases,'' or ``NSC-bases'' for short), with the understanding that the blown-up
smooth surfaces that support elliptic Calabi-Yau fibrations are no
longer strictly $\C^*$-surfaces.  One strong argument in favor of including
these bases in our analysis is that the base with the largest known
value of  $h^{1, 1} (B) = 491$, corresponding to the 6D F-theory
compactification with the largest gauge group, is of this type.  This
geometry was first identified in \cite{Candelas-pr,
Aspinwall-Morrison-instantons}, and was studied further in
\cite{toric, WT-Hodge}.  This geometry arises from a toric
base that contains two $-11$ curves that cannot be associated with
$D_{0, \infty}$, so the blown-up smooth base surface is neither
strictly toric nor $\C^*$.  We included toric bases with $-9, -10,
-11$ curves in \cite{toric}.  In the next section we discuss how
these are systematically included in the enumeration of $\C^*$-bases in the analysis of this paper.

For  (NSC) surfaces that contain -9, -10, or -11 curves, an additional
contribution to $T$ must be added for each blow-up needed to reach a
smooth surface with only allowed clusters ({\it i.e.,} with -12 curves
instead), so the equation for $T$ is modified from \eq{eq:t-equation}
to
\begin{equation}
T = h^{1, 1} (B) -1 =(\sum_{i = 1}^{N}  k_i) -N + 
c_{11} + 2c_{10} + 3c_{9} +
1 \,,
\label{eq:t-equation-general}
\end{equation}
where $c_k$ is the number of curves with self-intersection $-k$. 

\section{Enumeration of bases}
\label{sec:enumeration}

We now summarize the results of a full enumeration of the
set of $\C^*$-base surfaces relevant for F-theory
compactification.
In this section we describe the features relevant for the
corresponding 6D supergravity theories.  The following section
(\S\ref{sec:features}) focuses
on aspects of the associated Calabi-Yau geometries.

\subsection{Classification and enumeration of $\C^*$-bases}

Any $\C^*$-surface can be characterized by the self-intersection
numbers $n_0, n_\infty$ of the divisors $D_0, D_\infty$ associated
with the sections $\Sigma_\pm$, the number $N$ of
chains associated with distinct fibers 
along which blow-ups have occurred, and the intersection structure of the
divisors connected along each chain, characterized by the integers
$n_{i, j}$ as described above.  Thinking of each base $B$ as a blow-up
of $\F_m$, each chain is constructed by first blowing up a
point at the intersection of a fiber $F_i$ in $\F_m$ with either $D_0$
or $D_\infty$, giving a chain of length 2 containing a
pair of intersecting $-1$ curves. 
Additional blow-ups can then be performed at the intersections either between pairs of
adjacent divisors along the chain, or between the divisors at the end
of the chain and $D_0$ or $D_\infty$. Each nontrivial chain thus is
associated with a decrease in the self-intersection of $D_0$ or
$D_\infty$ by at least one.  Since the minimum values of $n_0,
n_\infty$ are $-12$ (as determined by the allowed non-Higgsable clusters), and their associated intersection numbers for $\Sigma_{\pm}$ on $\F_m$ are
equal and opposite, the maximum number of possible nontrivial chains
is $N = 24$.
Constructions with $N = 2$ or fewer nontrivial chains are realized
already in the toric context described in \cite{toric}; we have
included these cases in the complete analysis here, though the
counting is slightly different due to the treatment of $-9, -10,$ and
$-11$ curves.

To enumerate all possible $\C^*$-bases then, we can proceed by first
constructing all possible nontrivial chains associated with at most 24
blow-ups of points on the ending divisors of each chain (similar constructions were described in \cite{toric, Heckman-mv}).  We then
consider all possible combinations of these chains compatible with the
bound on intersection numbers of the sections $D_{0, \infty}$.  To
implement this enumeration algorithmically, we consider the set of
partitions of $k$ such that $1 \leq k \leq 24$, which are equivalent
to all valid combinations for the number of blow-ups at the
intersection of distinct fibers with $D_{0}$ or $D_{\infty}$ for any
$\F_{m}$. Then for each $\F_{m}$ and partition $\lambda \vdash k$ such
that $k=k_{1}+\ldots+k_{N}$, we can identify all combinations of
chains associated with $k_{1}, \ldots,k_{N}$ blow-ups on the ending
divisors such that the intersection of these chains with $D_{0}$ and
$D_{\infty}$ collectively satisfy the F-theory rules contained in
Table~\ref{t:clusters} and Table~\ref{t:NHCs}. Doing so for every
achievable combination of $n_{0}, n_{\infty}$ with $\lambda \vdash k$
gives a systematic method for enumerating all valid
$\C^*$-bases.

We have carried out the complete enumeration of all strictly
$\C^*$-bases ({\it i.e.,} including no -9, -10, -11 curves) for
all values of $N$ from 0 to 24.  We have also enumerated those bases
which are not strictly $\C^*$ (or toric), due to -9, -10, or -11
curves on the fiber chains.  We have not considered bases with -9, -10,
or -11 curves for $D_0$ or $D_\infty$, since blowing up these curves
gives a $\C^*$-base with larger $N$, so including these would simply
amount to overcounting.  The cases $N = 0, 1, 2$ represent toric
bases.  We have not included toric bases with -9, -10, -11 curves on
fiber chains when there is an equivalent toric base where such a curve
can be mapped to $D_0$ or $D_\infty$ by a rotation of the loop of
toric divisors.  These cases are already counted elsewhere in the set
of $\C^*$ bases, as discussed in more detail  in the following section. 

\subsection{Distribution of bases}
\label{sec:distribution}

The number of distinct bases, uniquely determined by $N, n_0,
n_\infty$, and the intersection configuration of the curves on the $N$
chains (up to the symmetries associated with permutations of the
chains and
the simultaneous reflection of all chains
combined with $n_0 \leftrightarrow n_\infty$), is tabulated for each $N$ from 0
to 24
in
Table~\ref{t:table}.  
Note that in cases where the base is toric and could have either $N =
1$ or $N = 2$ (this can occur when there is a curve of
self-intersection 0 that can either appear as a fiber or as one of $D_{0}, D_{\infty}$) we have counted the base as having the minimal value, $N
= 1$.
We find a total of 
126,469
smooth
$\C^*$-bases that are acceptable for 6D F-theory
compactifications. 
There are an additional 35,935 bases that come from $\C^*$-bases with -9, -10, and -11 curves in the fiber chains, for a total of
162,404 allowed bases.
We use this larger set for the statistical analyses in the remainder
of this paper.
The largest fraction
(nearly 38\%) of these bases come from the case with
3 non-trivial chains, and the number of
allowed bases decreases as the number of non-trivial chains increases
above 3.

\begin{table} 
\label{t:table}
\begin{center}
\begin{tabular}{| c | 
c |c | c| c| c|
}
\hline
\hline
\# Fibers $N$ &  $\C^*$-bases & $\C^*+$ NSC bases & Max $T$ & Min $T$ & Peak $T$
\\
\hline
$\P^2$ & 1 & 1 & 0 & 0 & 0\\
\hline
0 & 10 & 10 & 1 & 1 & 1\\
1 & 9,383 & 14,183 & 193 & 2 & 18 \\
2 & 25,474 & 28,733 & 182 & 3 & 21\\
3 & 44,930 & 61,329 & 171 & 4  & 21\\
4 & 20,980 & 27,134  & 160 & 5 & 21\\
5 & 11,027 & 13,811  & 149 & 6  & 21\\
6 & 6,137 & 7,462  & 138 & 7  & 25\\
7 & 3,485 &  4,133 & 127 & 8 & 25\\
8 & 2,034 & 2,356  &116 & 9 & 25\\
9 & 1,190 & 1,329  &105 & 10 & 25\\
10 & 709 & 768  & 94 & 11 & 25 \\
11 & 423 & 449  & 83 & 12  & 25\\
12 & 262 & 273  & 72 & 13 & 25 \\
13 & 159 & 164  & 61 & 14 & 25\\
14 & 101 & 104  & 61 & 15 & 25\\
15 & 62 & 63 & 39 & 16 & 25\\
16 & 40 & 40 & 30 & 17 & 25 \\
17 & 24 & 24 & 27 & 18 & 25\\
18 & 16 & 16 & 26 & 19 & 25\\
19 & 9 & 9 & 25 & 20 & 25 \\
20 & 6 & 6 & 25 & 21 & 25\\
21 & 3 & 3 &25 & 25  & 25\\
22 & 2 & 2 & 25 & 25 & 25 \\
23 & 1 & 1 & 25 & 25 & 25\\
24 & 1 &  1 & 25 & 25 & 25\\
\hline
total & 126,469 & 162,404 & 193 & 0& 25\\
\hline
\hline
\end{tabular}
\end{center}
\caption[x]{\footnotesize The number of distinct $\C^*$-bases, with
  two divisors $D_{0}$ and $D_{\infty}$ associated with sections
  $\Sigma_{\pm}$ connected by $N$ chains of curves of negative
  self-intersection.  Cases $N = 0, 1, 2$ correspond to toric bases.
  $P^2$ is listed separately.  Toric bases with a single 0-curve that
  can either be a section $(N = 2)$ or a fiber $(N = 1)$ are listed as
  $N = 1$ bases.  NSC refers to bases that are not strictly
  $\C^*$-surfaces in that they are described by $\C^*$-surfaces with
  -9, -10, or -11 curves in the fiber chains whose blow-up takes the
  base outside the set of $\C^*$-surfaces.  The bases described using
  toric geometry in \cite{toric} all appear in this table, though some
  are not strictly toric and appear in column 3 and/or in rows with $N
  > 2$ due to additional fibers produced when blowing up $-9, -10,
  -11$ curves.   Peak $T$ refers to the value of $T$ that occurs for
  the greatest number of bases at each $N$. Both NSC  and strictly $\C^*$-surfaces were used to determine the maximum, minimum, and peak $T$ values.}
\end{table}

As mentioned above, the tabulation of bases given in
Table~\ref{t:table} differs from that of \cite{toric} in the way that
toric bases with $-9, -10,$ and $-11$ curves are treated.  In
\cite{toric}, 61,539 bases were identified based on toric structures
that in some cases included $-9, -10,$ or $-11$ curves.  These bases
include the 34,868 strictly toric bases listed for $\P^2$ and $N=0,1$ and $2$
in column 2  of Table~\ref{t:table}, another 8,059 bases that arise
from toric bases with $-9, -10,$ or $-11$ curves on the fiber chains,
included in column 3 of the table, and another 18,612 bases that are
included in rows $N > 2$ and correspond to $\C^*$-bases (with or
without $-9, -10,$ or $-11$ curves on the fiber chains).  Bases in the
last category arise from
toric bases with $-9, -10, -11$ curves on the sections $D_0, D_\infty$
that give extra nontrivial fibers when the necessary points are blown
up.  To summarize, the analysis here includes all the bases
constructed in \cite{toric}, but discriminates more precisely based on
the detailed structure of these bases.

The parameter $T$ (number of tensor multiplets) serves as the most significant distinguishing
characteristic of 6D supergravity theories, and is related to the
simplest topological feature $h^{1, 1}(B)$ of the base surface $B$. Since each nontrivial fiber involves at least one blowup, and $T =1 $ for all $\F_{m}$, $T \geq N+1$. 
The value of $T= h^{1, 1}(B) -1$ for the $\C^*$-bases with $N > 2$
ranges from $T=4$ to $T = 171$.  The largest $T$ for any known base is
$T = 193$; this is an NSC base with only one nontrivial fiber ($N = 1$).
Heuristic arguments were given in \cite{toric} that no other 
base can have $T > 193$; as discussed in the next section this
conclusion is supported by the results of this paper.  
Note that for $N$ from 1 to 13, the maximum value of $T$ drops by
precisely 11 for each increment in $N$.  This can be understood from
the appearance of specific maximal sequences containing $-12$ 
curves, as discussed in \S\ref{sec:groups}.

\begin{figure}
\begin{center}
\includegraphics[width = 12.0cm]{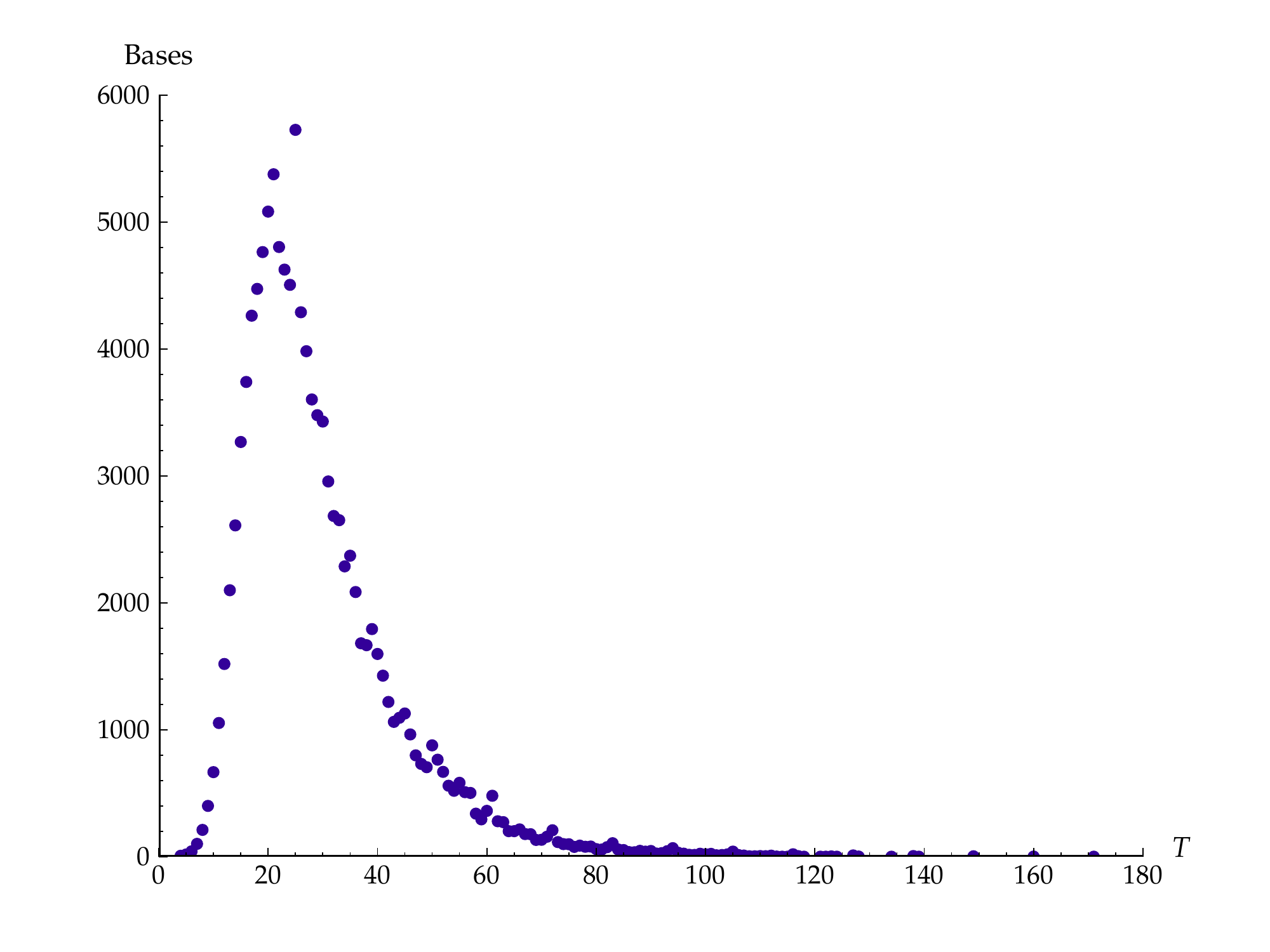}
\end{center}
\vspace{-1 cm}
\caption[x]{\footnotesize Number of
distinct $\C^*$-bases as a function of the number of tensor
multiplets $T$}
\label{f:t-plot}
\end{figure}

The number of different bases that appear at each $T$ is plotted in
Figure~\ref{f:t-plot}. The number of bases peaks at $T = 25$.
The distribution of bases also peaks at $T = 25$ for every specific $N > 5$, out to $N = 21,
22, 23$ and $24$, where all possible $\C^*$-bases have $T = 25$.  One
feature of many of the bases with $T = 25$ is that they contain
precisely two $(-12)$ clusters giving $\ge_8$ gauge summands, and no
other non-Higgsable clusters. The
primary difference between these bases is that they have different
numbers of nontrivial chains and contain different combinations of
intersecting $-2$ curves.  As mentioned above, clusters of $-2$ curves
without curves of lower self-intersection generally arise at special
points in moduli spaces of bases without such clusters.  Indeed, many
of the $T = 25$ bases with gauge algebra $\ge_8\oplus \ge_8$ are limits of the same complex geometry.  We discuss this issue further
in \S\ref{sec:redundancies}, in the context of the full Calabi-Yau
threefold geometry of the elliptic fibration over $B$.  In a similar
fashion, many bases have $T = 21$ and contain one $(-12)$ cluster and
a single $(-8)$ cluster associated with a $\ge_7$ algebra summand.

\begin{figure}
\begin{center}
\includegraphics[width = 12.0cm]{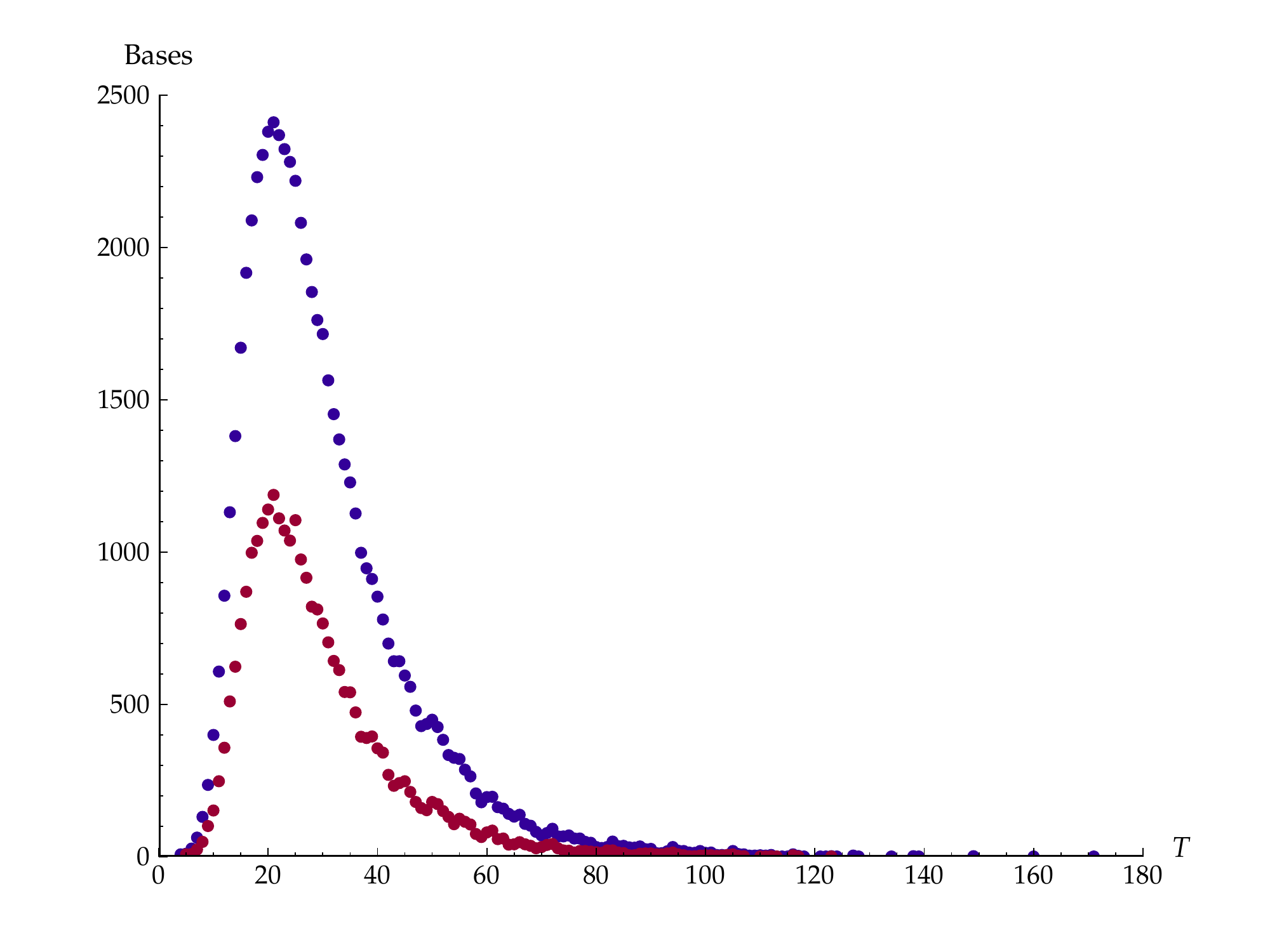}
\end{center}
\vspace{-1 cm}
\caption[x]{\footnotesize Number of distinct $\C^*$-bases with $N=3$ Fibers (upper blue data) and $N=4$ Fibers (lower purple data) for different numbers $T$ of tensor multiplets.}
\label{f:nt-plot}
\end{figure}

The number of $\C^*$-bases drops off rapidly between about $T = 30$
and $T = 60$.  Similar behavior for the toric subset of the bases was
noted in \cite{toric}.  In addition to the primary peak at $T = 25$,
there are smaller peaks in the distribution of bases starting at
$T=50$ and appearing  at intervals of eleven, so that they are visible at $T=61,\, 72 ,\, 83 , 94 ,\, $ {\it
etc.}  This feature is also visible when we consider the $N$-chain
cases separately for $3 \leq N \leq 14$ (see Figure~\ref{f:nt-plot}).
Again, the increment by 11 is related to $\ge_8$ chain sequences,
though the geometry seems less restricted as  $T$
increases.

It may seem surprising that the number of possible bases drops off so
rapidly with $N$.  Naively, one might imagine a set of $K$ relatively
short chains that could be combined arbitrarily in roughly $K^N/N!$
ways, which could grow rapidly with increasing $N$ for even modest
values of $K$.  The fact that the number of bases decreases rapidly
for increasing $N$ indicates that in fact, not many chains can be
combined arbitrarily.  These constraints come in part from the
intersection rules that drastically reduce the possibilities of which
chains can simultaneously intersect one of the sections, and in part
from the increased number of blow-ups at points on the sections that
are needed to build complicated chains.  The relatively controlled
dependence on $N$ of the number of models suggests that going beyond
the class of $\C^*$-surfaces  to completely generic base surfaces may also give
a reasonably controlled number of possible bases.

\subsection{Gauge groups and chain structure}
\label{sec:groups}

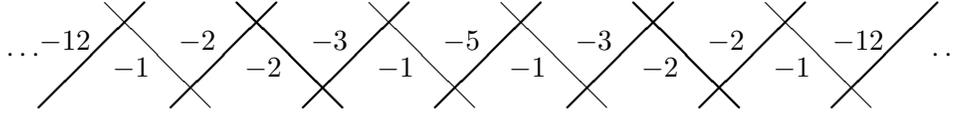
\begin{figure}
\begin{center}
\begin{picture}(200,100)(- 100,- 150)
\multiput(-145,-100)(50,0){6}{\line(1,-1){40}}
\thicklines
\multiput(-170,-140)(50,0){7}{\line(1,1){40}}
\multiput(-95,-100)(150,0){2}{\line(1,-1){40}}
\multiput(-160,-115)(300,0){2}{\makebox(0,0){$-12$}}
\multiput(-135,-125)(250,0){2}{\makebox(0,0){$-1$}}
\multiput(-110,-115)(200,0){2}{\makebox(0,0){$-2$}}
\multiput(-85,-125)(150,0){2}{\makebox(0,0){$-2$}}
\multiput(-60,-115)(100,0){2}{\makebox(0,0){$-3$}}
\multiput(-35,-125)(50,0){2}{\makebox(0,0){$-1$}}
\multiput(-10,-115)(300,0){1}{\makebox(0,0){$-5$}}
\put(-175,-120){\makebox(0,0){$\cdots$}}
\put(175,-120){\makebox(0,0){$\cdots$}}
\end{picture}
\end{center}
\caption[x]{\footnotesize  Bases having large values of $h^{11} (B) =
  T + 1$ have an intersection structure dominated by multiple
  repetitions of a characteristic sequence of intersecting divisors
  associated with non-Higgsable gauge algebra $\ge_8 \oplus\gf_4
\oplus 2 (\gg_2 \oplus\gsu (2))$.}
\label{f:e8-sequence}
\end{figure}

\begin{figure}
\begin{center}
\includegraphics[width = 12.0cm]{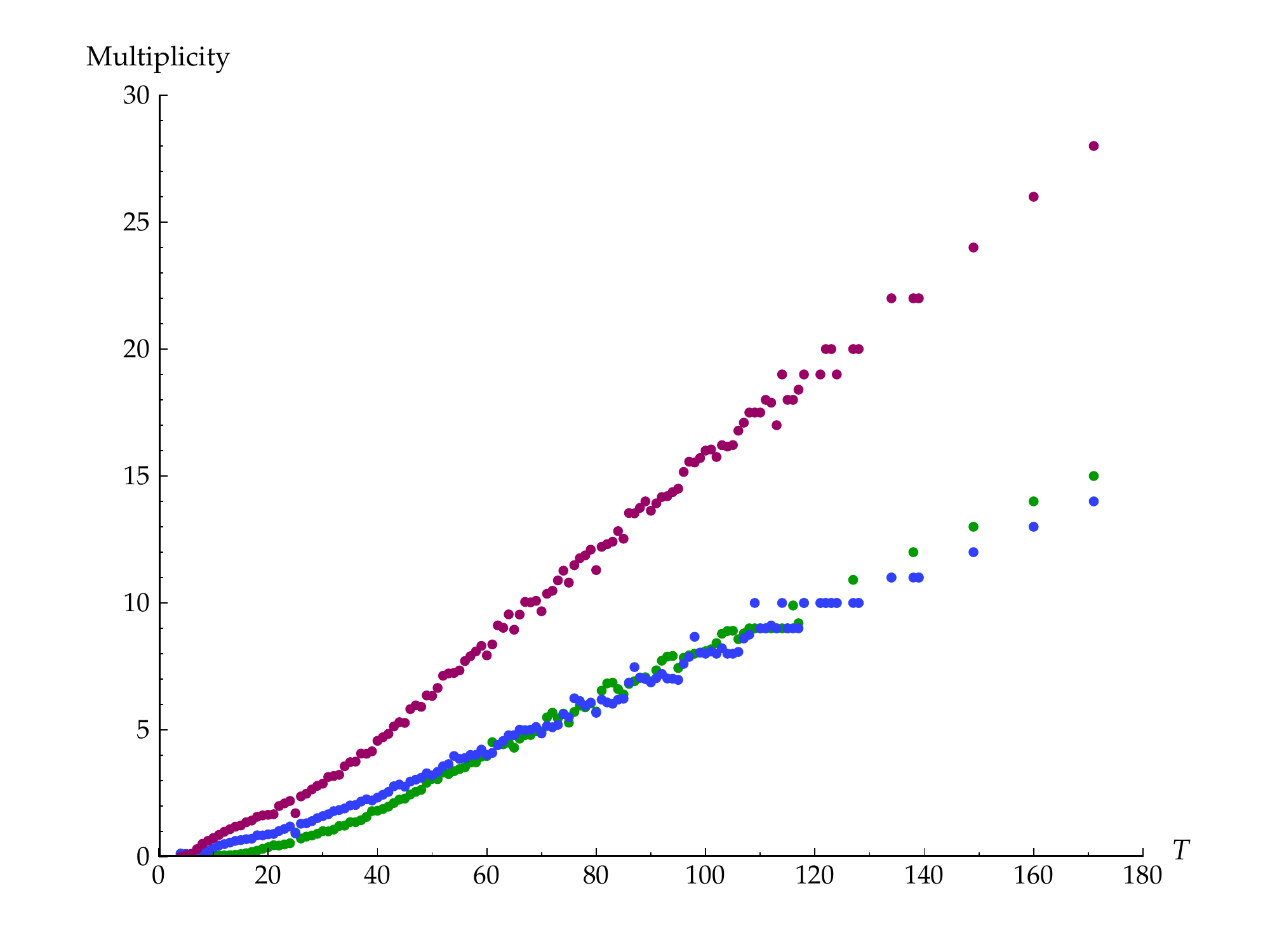}
\end{center}
\vspace{-1 cm}
%\caption[x]{\footnotesize Average number of gauge algebra summands of
%  types $(\gg_2 \oplus\gsu (2))$, $\gf_4,$ and $\ge_8$ as a function of $T$.}
\caption[x]{\footnotesize Average number of gauge algebra summands as a function of $T$ for summands (ordered from top at $T=171$)  $\gg_2 \oplus\gsu (2)$, $\ge_8,$ and $\gf_4$.}
\label{f:gauge-factors}
\end{figure}

We have investigated a number of aspects of the class of $\C^*$-bases that we have identified.  One of the main conclusions of this
investigation is that the structure of the $\C^*$-bases with large
$T$ is very similar to that of toric bases with large $T$.  For toric
bases with large $T$, the intersection structure of divisors with
negative self-intersection is largely based on long chains dominated
by sequences of maximal units with gauge algebra $\ge_8 \oplus\gf_4
\oplus 2 (\gg_2 \oplus\gsu (2))$ (See Figure~\ref{f:e8-sequence}). 
The same is true of $\C^*$-bases.
The number of gauge  algebra summands of these types are plotted in
Figure~\ref{f:gauge-factors} against $T$, and grow linearly in a
fashion very similar to that in the toric case.
The linear sequences of 12 curves giving the gauge algebra
contributions 
 $\ge_8 \oplus\gf_4
\oplus 2 (\gg_2 \oplus\gsu (2))$
were
described in \cite{clusters}; we refer to these as ``$\ge_8$ units''
for simplicity, as they appear frequently in the intersection
structure of bases with large $T$. These units are maximal in the sense that they can not be blown up in any way consistent with the existence of an elliptic fibration.

\begin{figure}
\begin{center}
\includegraphics[width=12cm]{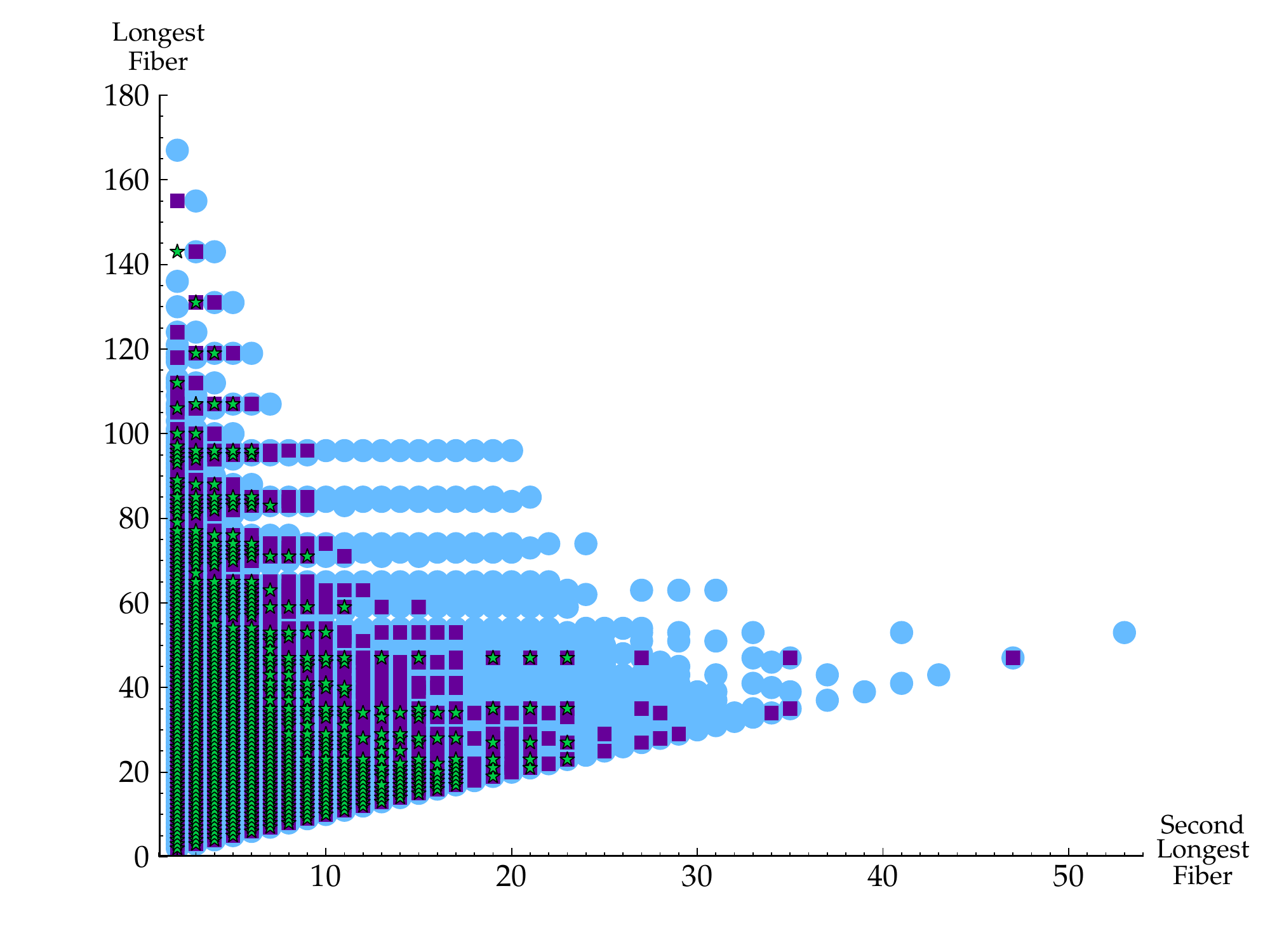}
\end{center}
\caption[x]{\footnotesize  Comparison of the lengths of the longest
  and second longest chains associated with distinct fibers for bases with $N = 3$ (blue circles), $N = 4$ (purple squares), and $N=5$ (green stars). }
\label{f:chains-34}
\end{figure}

A detailed analysis of the configurations at large $T$ makes this
correspondence clear.  
As discussed in \cite{clusters, toric}, the base with  $T = 193$ is a (NSC) base with $N = 1$, where the single fiber chain is essentially
16 $\ge_8$ units ending in $-12$ curves at $D_0, D_\infty$; the two
$-12$ curves one $\ge_8$ unit away from the ends of the fiber are actually $-11$
curves in the toric base, which is what makes this base not strictly toric or
$\C^*$ (and therefore places it in the third column of Table~\ref{t:table}).
The $N = 2$ base with $T = 182$ is identical except that the long
fiber chain contains only 15 $\ge_8$ units (again with $-11$
curves one $\ge_8$ unit away from each end), and the second chain is a simple $(-1,
-1)$ chain arising from a single blow-up on one of the sections.  The
unique $N = 3$ $\C^*$-base with $T
= 171$ is again similar, with $n_0 = n_\infty = -12,$, two $(-1, -1)$
chains, and the longer chain being a sequence of 14 $\ge_8$ units, again
with two $-11$ curves in the $\C^*$-base.
Other bases with large $T$ are dominated in a
similar fashion by one or two very long chains, with the other chains
generally being of type $(-1, -1)$ or other very short chains.
For example, there is only one base with $N = 3$ that has two fiber
chains
  of length $> 50$ (see Figure~\ref{f:chains-34}).  For this base
  there are two long chains of length 53 that combine to form a loop
  of 9 $e_8$ chains, with a $-12$ curve and the opposite $-5$ curve
  for $D_0$ and $D_\infty$ (and the usual pair of $-11$ curves near
  the end on each side), and the third fiber chain is of the form
  $(-1, -1)$.  Over all $N = 3$ bases, when the shortest fiber chain has
  length $> 3$ then the next shortest is of length $23$ or less.

This analysis of large $T$ $\C^*$-bases strengthens the conclusion
argued heuristically in \cite{toric} that no base can
have $T > 193$.  In particular, the $\C^*$-bases allow two new types
of topological structure to the intersection diagram: loops and
branches.  Neither of these types of structure allows for new
constructions that qualitatively change the nature of the bases at
large $T$.  In general, bases with large $T$ have a long linear
sequence of $\ge_8$ chains, with additional loops rapidly decreasing
the maximum value of $T$.  There are no new features associated with
branches that suggest any mechanism for constructing bases with large
$T$ and more complicated intersection structures.  
In fact, the known base with the largest $T$ corresponds to simply a
single long $\ge_8$ chain, with no branching or loops at all.  The
only way to modify this structure topologically is to add branchings
and loops, and the results that we have found here show that adding up
to two branchings and an arbitrary number of loops does not give any
way of increasing $T$; rather, as the topological complexity increases
the upper bound on possible $T$'s  decreases.
Thus,
while we still do
not have a proof that $T = 193$ is the maximum possible for a smooth
base $B$, the absence of new structure in the $\C^*$-base models seems
to make it unlikely that generic bases with even more complicated
branching and looping structures could have larger values of $T$,
even without the $\C^*$ restriction.

\begin{figure}
\begin{center}
\includegraphics[width=12cm]{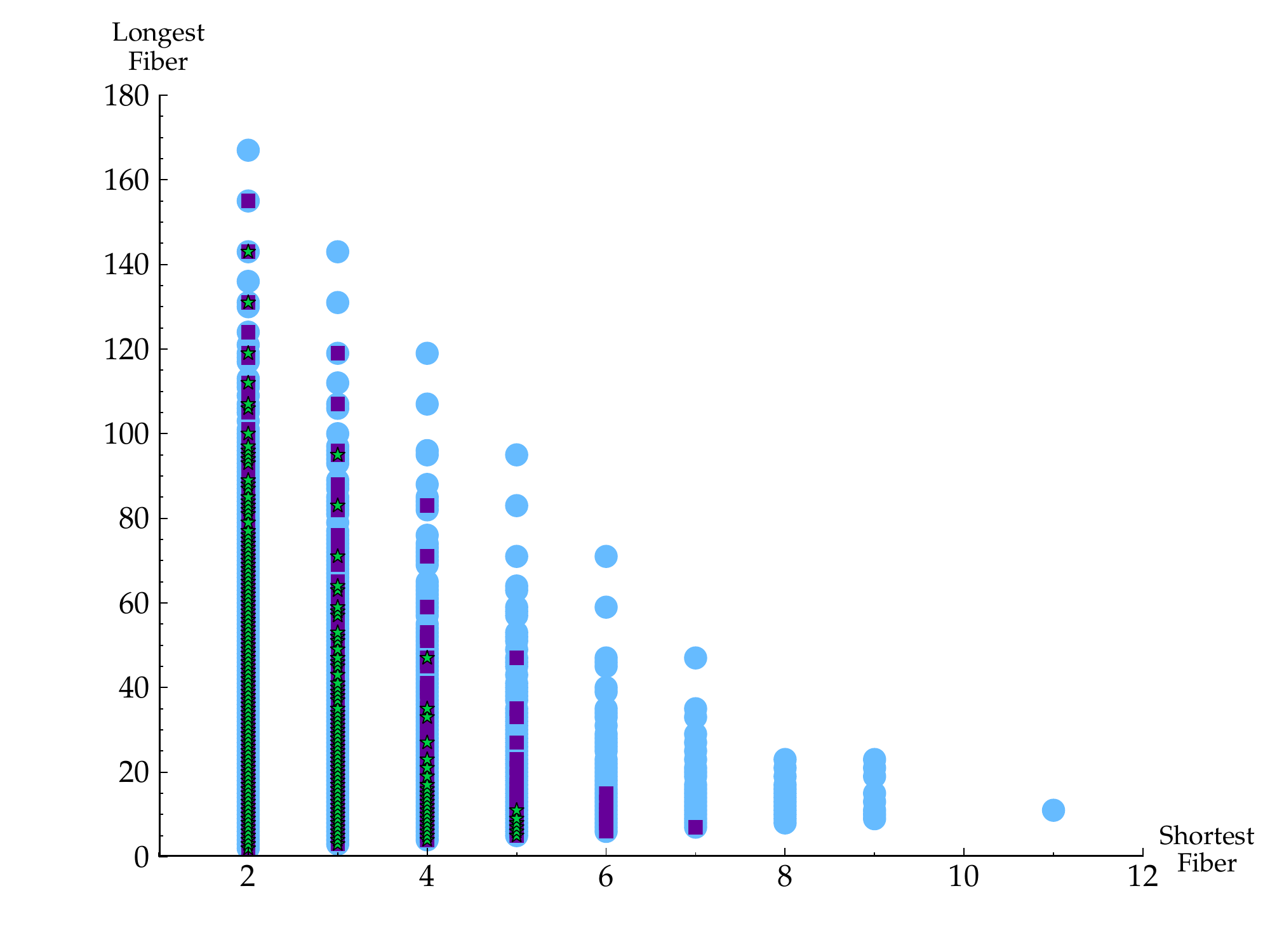}
\end{center}
\caption[x]{\footnotesize  Comparison of the lengths of the longest
  and shortest chains associated with distinct fibers for bases with $N = 3$ (blue circles), $N = 4$ (purple squares), and $N=5$ (green stars). }
\label{f:chains-34-b}
\end{figure}

\section{Calabi-Yau geometry}
\label{sec:features}

Over each F-theory base it is possible to construct an elliptically
fibered Calabi-Yau geometry.  Resolving singularities associated with
the nonabelian groups living on the discriminant locus gives a smooth
Calabi-Yau threefold.  While tuning the moduli over any base can
increase the degree of vanishing of the discriminant locus and hence
enhance the nonabelian gauge group, which corresponds to a change in the
associated smooth Calabi-Yau threefold, we focus here
primarily on the simplest
threefold over each base, corresponding to the maximally Higgsed 6D
supergravity model.

Each Calabi-Yau threefold $X$ has topological invariants given by the
Hodge numbers $\ho (X),\, \htt (X)$.  A large class of Calabi-Yau
threefolds that are realized as hypersurfaces in toric varieties were
identified by Kreuzer and Skarke \cite{Kreuzer-Skarke}, who produced a
comprehensive database of 474 million Calabi-Yau constructions of this
type.  These examples include threefolds with 30,108 distinct pairs of
Hodge numbers. The Hodge numbers associated with allowed toric bases
(including some NSC bases) were computed in \cite{WT-Hodge},
and it was shown that
the distinctive upper boundary of the ``shield region'' spanned by the
Kreuzer and Skarke Hodge data is associated with a trajectory of
blow-ups of the Hirzebruch surface $\F_{12}$.  In particular, it was
proven that the maximum possible value of $\htt (X)$ for \emph{any}
elliptically-fibered Calabi-Yau threefold (with section) is given by
\begin{equation}
\htt (X) \leq 491 \,,
% \label{eq:}
\end{equation}
independent of whether the base is toric or not.  In this section we
carry out an analysis of the Hodge numbers for the more general class
of $\C^*$-bases, allowing us to identify some new Calabi-Yau
threefolds with 
interesting features.

\subsection{Calabi-Yau geometry of elliptic fibrations over $\C^*$-bases}

The Hodge numbers of the smooth Calabi-Yau threefold $X$
associated with a generic elliptic fibration over the base $B$ are
related to the intersection structure of the base and to the gauge
group and matter content of the associated 6D supergravity theory.

The Hodge number $h^{1, 1} (X)$ is related to the structure of the
base by the Shioda-Tate-Wazir formula \cite{stw}, which in the
language of the 6D supergravity theory states \cite{Morrison-Vafa}
\begin{equation}
h^{1, 1} (X) = h^{1, 1} (B) + {\rm rank} (G) + 1
= T + 2 + {\rm rank} (G_{\rm nonabelian}) + V_{\rm abelian} \,,
\label{eq:h11}
\end{equation}
where $G$ is the full (abelian + nonabelian) gauge group of the 6D
theory.  While the nonabelian group is determined as described above
from the singularity structure of the discriminant locus, the rank of
the abelian group corresponds to the rank of the
\emph{Mordell-Weil} group of sections of the fibration.  
In general, the rank of the Mordell-Weil group is difficult to compute
mathematically (see, {\it e.g.}, \cite{Hulek-Kloosterman}).
The
Mordell-Weil group is a global feature of the elliptic fibration,
corresponding in the physics context to the number of
$U(1)$ factors in the 6D supergravity theory; the global aspect of 
this structure is what makes computation of the group a particularly
challenging problem.  Recent
progress on understanding abelian factors in general F-theory
constructions was made in 
\cite{Grimm-Weigand,
Park-Taylor, Morrison-Park, Park-abelian, Mayrhofer:2012zy,
  Braun:2013yti, bmpw, Cvetic-Klevers, Braun:2013nqa, bmpw-2, Cvetic-Klevers-2, Braun-fate, DPS, 
mt-sections}.

The Hodge number $h^{2, 1} (X)$ gives the number of
complex structure moduli of the threefold $X$.
This set of moduli represents all but one of the neutral scalar fields
in the 6D gravity theory -- the remaining scalar field is associated
with the overall K\"ahler modulus on the base $B$.
\begin{equation}
h^{2, 1} (X) = H_{\rm neutral} -1 \,.
% \label{eq:}
\end{equation}
The number of neutral hypermultiplets in the theory on a $\C^*$-base $B$ can be computed from the intersection structure on $B$
by analyzing the generic Weierstrass form
over that base, as we describe below in Section~\ref{sec:neutral}.  The
total number of scalar fields is also related through
the 6D gravitational anomaly cancellation condition to the number of
vector multiplets $V$
in the theory  \cite{gsw-6, Sagnotti}
\begin{equation}
H_{\rm neutral} + H_{\rm charged}-V = 273-29T \,.
\label{eq:anomaly}
\end{equation}
Since the number of charged hypermultiplets can be computed by adding
the contributions from Table~\ref{t:clusters} for each non-Higgsable
cluster in $B$, and the numbers of neutral hypermultiplets, tensors,
and nonabelian vectors are also computable from the intersection
data on $B$, this gives us a way of computing the total number of
abelian vector multiplets $V_{\rm abelian}$.  Using this in
\eq{eq:h11}, we can thus compute both Hodge numbers $\ho (X),\, \htt (X)$ in a
systematic way for any $\C^*$-base.

\subsection{Counting neutral hypermultiplets}
\label{sec:neutral}

Computing the number of neutral hypermultiplets over a given base $B$
can be done in a systematic fashion by following the sequence of
blow-ups needed to reach $B$ starting from a Hirzebruch surface
$\F_m$.  
In the case where $B$
is a toric surface, this analysis is particularly
simple and was described in \cite{toric}.
For a toric base, the number of neutral hypers is related to
the number of free parameters $W$ in the Weierstrass model for $B$
through
\begin{equation}
H_{\rm  neutral} = W-w_{\rm aut} + N_{-2}  \;\;\;\;\;
{\rm (toric)}\,,
\label{eq:hyper-counting}
\end{equation}
where $w_{\rm aut}$ is the dimension of the automorphism group of $B$,
and $N_{\rm -2}$ is the number of curves of self-intersection $-2$ in
$B$ that do not live in clusters carrying a gauge group.  Basically,
automorphisms of $B$ correspond to Weierstrass monomials that do not
represent physical degrees of freedom, while $-2$ curves represent
moduli that have been tuned to a special point, as discussed above.
As shown in \cite{toric}, each curve of self-intersection $k \geq
0$ in the
toric fan contributes $k +1$ to the dimension of the automorphism
group, with two additional universal
automorphisms corresponding to the toric
structure.  For toric surfaces, the number of Weierstrass monomials in
$f$ is simply given by the set of points $m$ in the lattice dual to
the toric fan that satisfy $\langle m, v\rangle \geq -4$ for all $v$
in the set of rays generating the fan, and a similar condition with
$\langle m, v\rangle \geq -6$ for monomials in $g$.  As discussed in
\cite{toric}, there is a simple geometric picture of this in the toric
context.

More generally, we can start on $\F_m$ with a given number of
monomials $W$ in the Weierstrass degrees of freedom of $f, g$ and
explicitly tune the moduli in $f, g$ to blow up the base at the
desired set of points to realize any base $B$.  We can then compute
the number of remaining Weierstrass moduli and apply
\eq{eq:hyper-counting}
(with one slight modification in the counting
of $N_{-2}$ as discussed further below).  This procedure can be
carried out in a clear fashion when the base $B$ is
$\C^*$.  The simplification in the $\C^*$ case comes from the
fact that we can treat each chain arising from a blown-up fiber on
$\F_m$ in a parallel fashion to the toric case.  In the toric case we
are blowing up points on only two fibers, which we can take to be at
the points $z = 0, \infty$ in a local toric
coordinate chart $(z, w) \in\C \times\C$ as discussed in Section
\ref{sec:toric}.  Consider the constraints on Weierstrass monomials
coming from blow-ups along the fiber at $z = 0$.  Each such constraint
corresponds to imposing a vanishing condition on a set of monomial
coefficients in an expansion $f = \sum_{n, m}c_{n,m}z^nw^m$, and similarly
for $g$.  The set of constraints imposed depends upon the sequence of
blow-ups in the fiber, and has a convenient geometric description in
the language of the toric fan and dual space.  For a $\C^*$-base
with $N$ chains located at $z = z_1, \ldots, z_N$, we simply impose
the appropriate set of constraints on the monomial coefficients in an
expansion $f = \sum_{n, m}c^{(i)}_{n,m} (z-z_i)^nw^m$ for the $i$th
chain.  This
gives a set of linear conditions on the coefficients in $f, g$.  
The
dimension of the space of independent solutions to these linear
constraints can then be used to compute the number of independent
Weierstrass monomials $W$, in an analogous fashion to
(\ref{eq:hyper-counting}).  As in the toric case, when one of the
sections $D_0, D_\infty$ has self-intersection $k \geq 0$ it
contributes $k +1$ to the dimension of the automorphism group.  For a
$\C^*$-base with $N = 3$ or more blown up fibers,
the first two fibers can be fixed at $z = 0, \infty$, using up two
automorphisms originally associated with the fibers of
self-intersection 0 in the toric base.
The blowing up of
the third fiber corresponds to fixing another point in $\C^*$, which can for
example be chosen to be $z = 1$, fixing one of the remaining
automorphisms and leaving only one of the universal two present in all
toric  bases.
For each of the $N -3$ fibers beyond the third, blowing up an
additional point involves choosing a value $z_i$ which is itself a
Weierstrass parameter though it does not correspond to a monomial.
Thus, the formula (\ref{eq:hyper-counting}) must be augmented by $N
-3$ when more than 3 fibers are blown up.  The complete formula for
the number of neutral hypermultiplets is given at the end of this
section, in \eq{eq:hyper-counting-general}.

\begin{figure}
\begin{center}
\centering
\begin{picture}(200,130)(- 100,- 70)
\put(-105, 0){\makebox(0,0){\includegraphics[width=8cm]{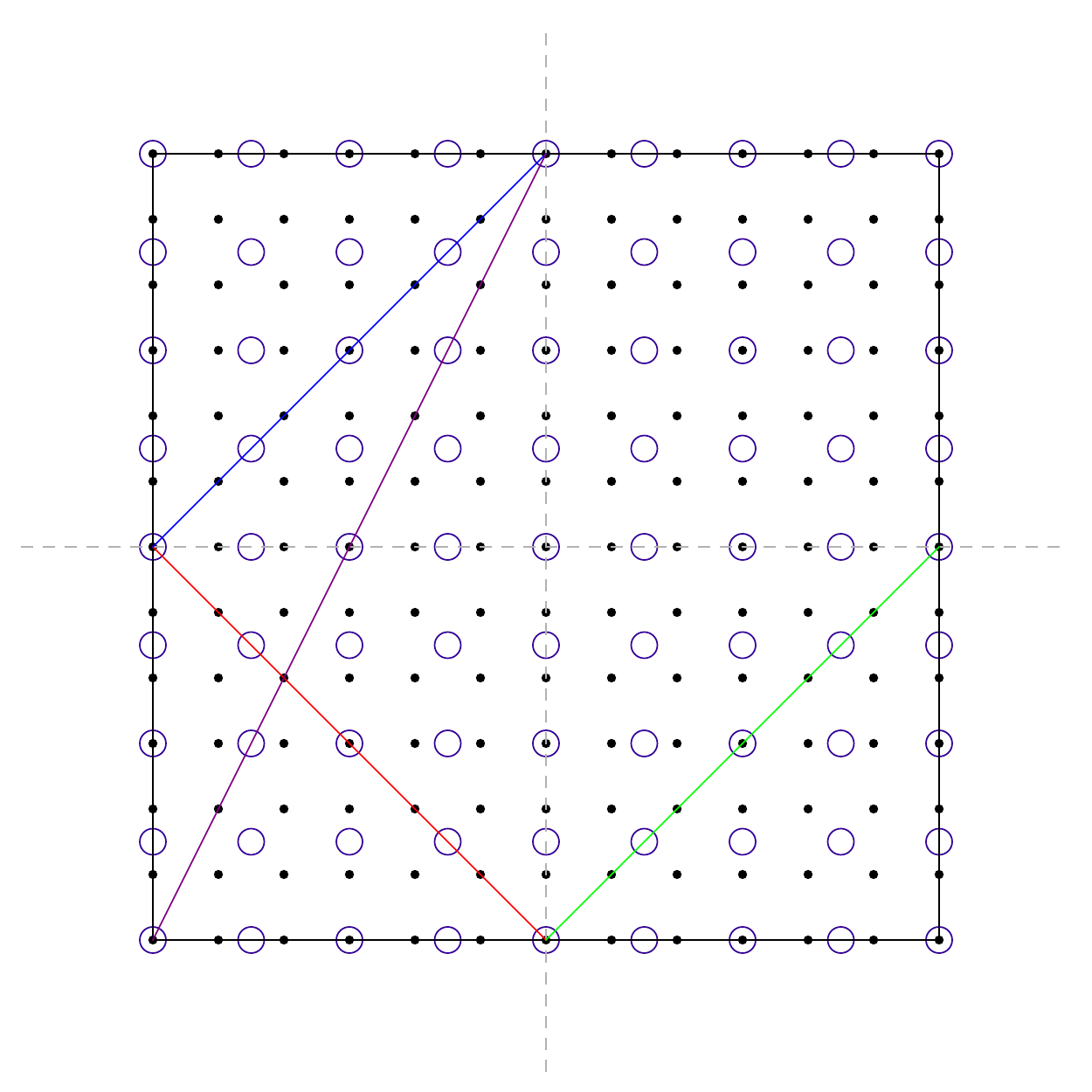}}}
\put(105, 0){\makebox(0,0){\includegraphics[width=5.5cm]{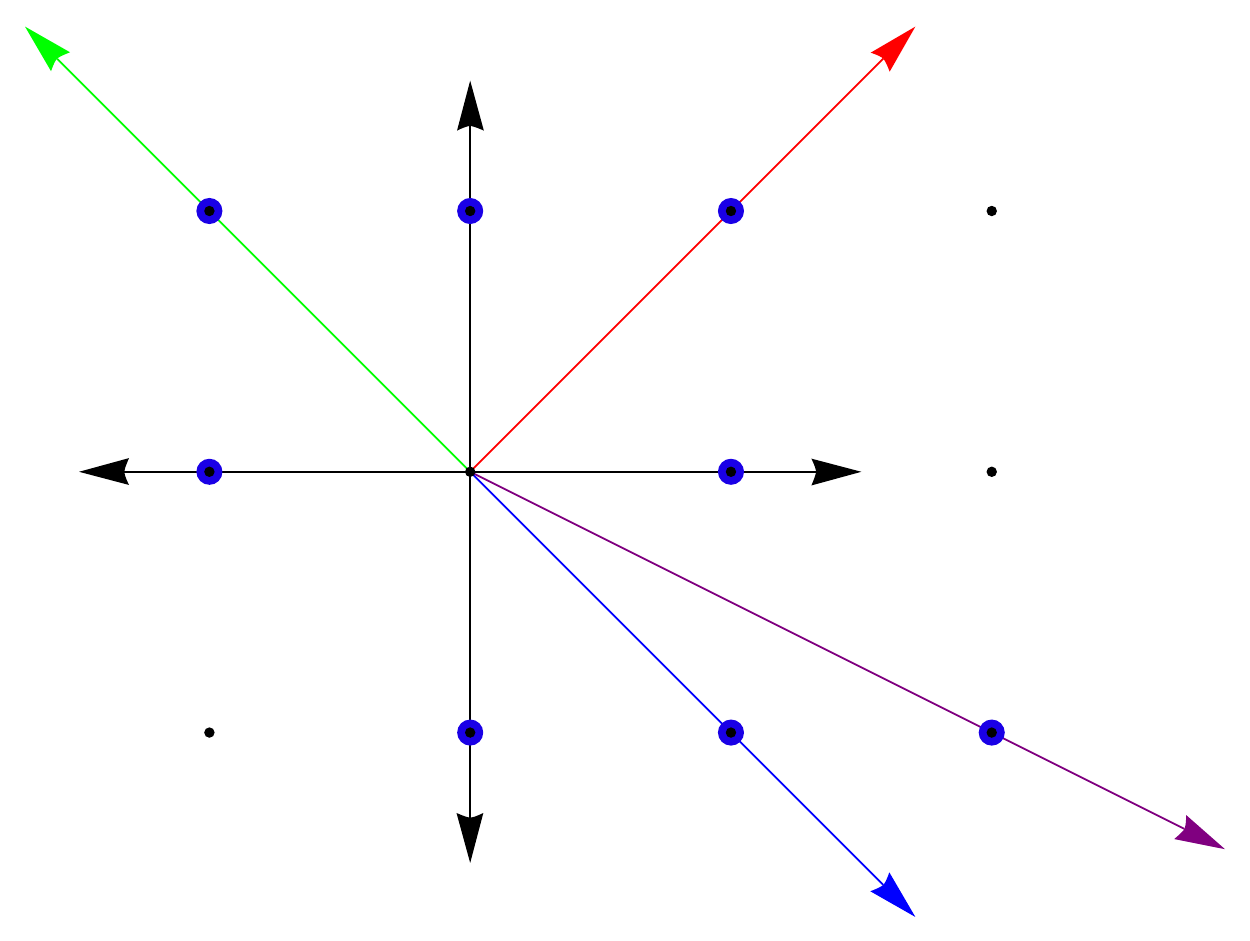}}}
\end{picture}
\end{center}
\caption[x]{\footnotesize An example of counting monomials in a toric
  base $B$.  The base $\F_0$ has 250 Weierstrass coefficients: $9 \times 9
  = 81$ in $f$ (solid dots) and $13 \times 13$ in $g$ (round circles)
  in the dual lattice.  These are all of the dots and circles in the
  diagram on the left.
Each blow-up corresponds to adding a vector to
  the fan (above right), which removes some monomials from $f$ and
  $g$.  For example, blowing up the point corresponding to the red ray
  in the diagram on the right removes all monomials below and to the
  left of the red line in the left-hand diagram, blowing up on the blue
ray removes all points above and to the left of the blue line,  {\it etc.}.
The base $B$ has 136 monomials, 44 in $f$ and 92 in $g$, corresponding
to the dots and circles in the left-hand diagram that lie above the red
and green lines and to the right of the purple line.}
\label{f:example-monomials}
\end{figure}

It may be helpful to illustrate this method with an example.
Consider first the toric base $B$ with the following parameters
\begin{eqnarray}
N = 2, &  & \; n_0 = -2, n_\infty = -1\\
{\rm chain}\ 1: &  &  (-1, -3, -1, -2)\\
{\rm chain}\ 2: &  &  (-1, -1)
\end{eqnarray}
This base has one $(-3)$ cluster, so the nonabelian gauge algebra
is $\gsu (3)$.
This base can be described as $\F_0$ with four points blown up, three
on the fiber $z = 0$ and one on the fiber at $z = \infty$.  A graphic
description of the toric fan and monomials in the dual lattice is
given in Figure~\ref{f:example-monomials}.

The monomials in this toric model can be computed using the methods of
\cite{toric}. The monomials in $f$ in the Weierstrass model on
$\F_0$ are of the form $c_{n, m}z^nw^m, 0 \leq n, m \leq 8$, and
similarly for $g= \sum_{n, m = 0}^{12}d_{n, m}z^nw^m$.  Blowing up the
points on $\F_0$ imposes conditions on the coefficients $c, d$.  For
example, blowing up the point $z = w = 0$ (corresponding to adding the
ray in red in the diagram on the right-hand side of Figure~\ref{f:example-monomials})
imposes the conditions that
$c_{n, m} = 0$ for all $n + m < 4$ and $d_{n, m} = 0$ for $n + m < 6$.
    This removes 31 Weierstrass moduli (and reduces the automorphism
    group by 2 by turning two $0$-curves into $-1$-curves), giving a
    change in the number of neutral hypermultiplets $\Delta H_{\rm
      neutral} = -29$, matching  \eq{eq:anomaly} with $\Delta T = 1$.
This corresponds to removing all the monomials on the lower-left
corner of the left-hand diagram in Figure~\ref{f:example-monomials}.    The analogous constraints imposed by blowing 
    up the other points on the fiber $z = 0$ as well as the point $z =
    \infty, w = 0$ are depicted in an analogous fashion in the Figure.
    Together, these constraints reduce the number of Weierstrass
    monomials to $W = 136$.  This matches with \eq{eq:hyper-counting}
    and \eq{eq:anomaly}, with $N_{-2}= w_{\rm aut} = 2$ and $V= 8$
    vector multiplets from the $\gsu(3)$ factor in the gauge algebra.
    Thus, we have determined the Hodge numbers of the resolved generic
    elliptic fibration over $B$
\begin{equation}
\ho (X) =   9,  \;\;\;\;\;\htt (X) =135 \,.
% \label{eq:}
\end{equation}

\begin{figure}
\setlength{\unitlength}{.83pt}
\begin{center}
\begin{picture}(180,160)(- 100,- 80)
\put(-35, 0){\vector(1,0){25}}
\put(-265, 60){\line(1, 0){225}}
\put(-265, -60){\line(1, 0){225}}
%\put(-155,70){\makebox(0,0){-1}}
%\put(-155,-70){\makebox(0,0){-2}}
\put(-155,70){\makebox(0,0){-2}}
\put(-155,-70){\makebox(0,0){-1}}
\put(-55,-70){\makebox(0,0){{\large$D_{\infty}$}}}
\put(-55,70){\makebox(0,0){{\large$D_{0}$}}}
% Fiber zero
\put(-255, 62){\line( -1, -3){12}}
\put(-255, -2){\line(-1, 3){12}}
\put(-255, 2){\line( -1, -3){12}}
\put(-255, -62){\line( -1, 3){12}}
\put(-277,14){\makebox(0,0){{\large$D_{1,2}$}}}
%\put(-250,14){\makebox(0,0){-1}}
\put(-250,14){\makebox(0,0){-3}}
\put(-277,48){\makebox(0,0){{\large$D_{1,1}$}}}
%\put(-250,48){\makebox(0,0){-2}}
\put(-250,48){\makebox(0,0){-1}}
\put(-277,-14){\makebox(0,0){{\large$D_{1,3}$}}}
%\put(-250,-14){\makebox(0,0){-3}}
\put(-250,-14){\makebox(0,0){-1}}
\put(-277,-48){\makebox(0,0){{\large$D_{1,4}$}}}
%\put(-250,-48){\makebox(0,0){-1}}
\put(-250,-48){\makebox(0,0){-2}}

% Fiber five
\put(-165, 5){\line( 1, -3){23}}
\put(-165, -5){\line(1, 3){23}}
\put(-170,35){\makebox(0,0){{\large$D_{2,1}$}}}
\put(-140,33){\makebox(0,0){-1}}
\put(-170,-35){\makebox(0,0){{\large$D_{2,2}$}}}
\put(-140,-33){\makebox(0,0){-1}}

% Fiber eleven
\put(-45, 62){\line( 0, -1){128}}
\put(-50,0){\makebox(0,0){0}}

%\put(-45,-60){\makebox(0,0){\includegraphics[width=0.4cm]{circle.eps}}}
\put(-45,60){\makebox(0,0){\includegraphics[width=0.4cm]{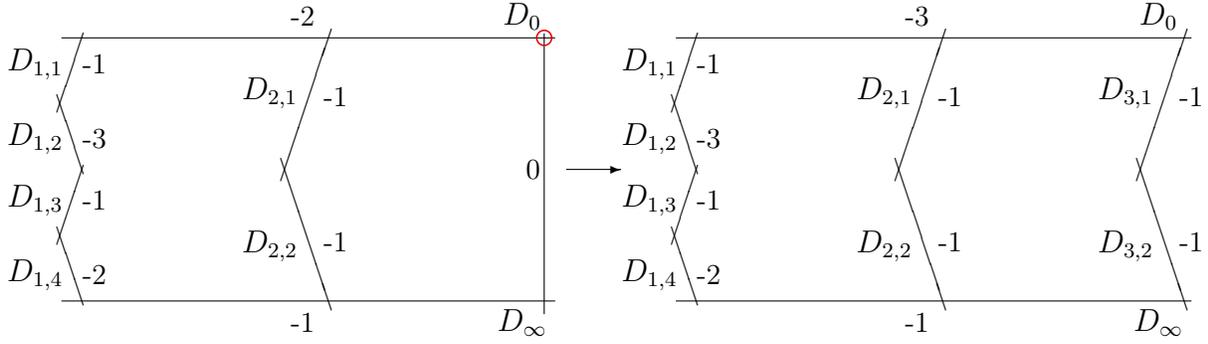}}}

% Second half
\put(15, 60){\line(1, 0){235}}
\put(15, -60){\line(1, 0){235}}
\put(125,70){\makebox(0,0){-3}}
\put(125,-70){\makebox(0,0){-1}}% WT: switched these too.
\put(235,-70){\makebox(0,0){{\large$D_{\infty}$}}}
\put(235,70){\makebox(0,0){{\large$D_{0}$}}}
% Fiber one
\put(25, 62){\line( -1, -3){12}}
\put(25, -2){\line(-1, 3){12}}
\put(25, 2){\line( -1, -3){12}}
\put(25, -62){\line( -1, 3){12}}
\put(3,14){\makebox(0,0){{\large$D_{1,2}$}}}
%\put(30,14){\makebox(0,0){-1}}
\put(30,14){\makebox(0,0){-3}}
\put(3,48){\makebox(0,0){{\large$D_{1,1}$}}}
%\put(30,48){\makebox(0,0){-2}}
\put(30,48){\makebox(0,0){-1}}
\put(3,-14){\makebox(0,0){{\large$D_{1,3}$}}}
%\put(30,-14){\makebox(0,0){-3}}
\put(30,-14){\makebox(0,0){-1}}
\put(3,-48){\makebox(0,0){{\large$D_{1,4}$}}}
%\put(30,-48){\makebox(0,0){-1}}
\put(30,-48){\makebox(0,0){-2}}

% Fiber six

\put(115, 5){\line( 1, -3){23}}
\put(115, -5){\line(1, 3){23}}
\put(110,35){\makebox(0,0){{\large$D_{2,1}$}}}
\put(140,33){\makebox(0,0){-1}}
\put(110,-35){\makebox(0,0){{\large$D_{2,2}$}}}
\put(140,-33){\makebox(0,0){-1}}
% Fiber twelve
\put(225, 5){\line( 1, -3){23}}
\put(225, -5){\line(1, 3){23}}
\put(220,35){\makebox(0,0){{\large$D_{3,1}$}}}
\put(250,33){\makebox(0,0){-1}}
\put(220,-35){\makebox(0,0){{\large$D_{3,2}$}}}
\put(250,-33){\makebox(0,0){-1}}
\end{picture}
\end{center}
\caption[x]{\footnotesize  An example of a $\C^*$-surface $B'$
given by
blowing up the toric surface $B$ from Figure~\ref{f:example-monomials}
at a
point.}
\label{f:st-example}
\end{figure}

Now, consider blowing up a point on a third fiber at $z = 1$ to form
the $\C^*$-base $B'$ shown in Figure~\ref{f:st-example}.  Blowing up at
the point $(z, w) = (1, 0)$ imposes the condition that $f$ and $g$
must vanish to degrees 4 and 6 in $(z-1)$ and $w$.  Consider for
example $f$.  From the geometry depicted in
Figure~\ref{f:example-monomials} it is clear that the only
undetermined coefficients of $f$ at leading orders in $w$ are
\begin{equation}
f = c_{4, 0} z^4 + c_{3, 1} z^3w + c_{4, 1} z^4w + c_{5,1} z^5w
+ \sum_{n = 2}^{6} c_{n,2} z^nw^2 + \cdots \,.
% \label{eq:}
\end{equation}
The condition that this vanishes to degree 4 in $(z-1), w$ 
forces all coefficients $c_{n, 0}$ and $c_{n,1}$ to vanish, imposes
two constraints at $m = 2$, and one constraint at $m = 3$, totaling 7 new constraints.  Similarly, coefficients of $g$ experience
14 further constraints, for a total of 21 constraints.  This reduces
the number of Weierstrass monomials to $W = 115$. The equations
\eq{eq:hyper-counting} and \eq{eq:anomaly} still apply, where now
$w_{\rm aut} = 1$ since one of the toric automorphisms is broken by
the reduction to $\C^*$ structure, and $N_{-2} = 1$ because the
blow-up at $z = 1, w = 0$ changes $n_0$ from $-2$ to $-3$.
With one new $-3$ curve the number of vector multiplets becomes $16$,
and the resolved generic elliptically fibered Calabi-Yau $X'$ over $B'$ has
Hodge numbers
\begin{equation}
\ho (X') = 12, \;\;\;\;\;\htt (X') = 114 \,.
% \label{eq:}
\end{equation}

In this way we can determine the Hodge numbers of the Calabi-Yau
threefolds associated with generic elliptic fibrations over
all $\C^*$-bases.  When the additional fibers added are more
complicated, the conditions on $f, g$ at  $z = z_i$ can be determined
by simply translating the conditions at  $z = 0$ from the toric
picture.

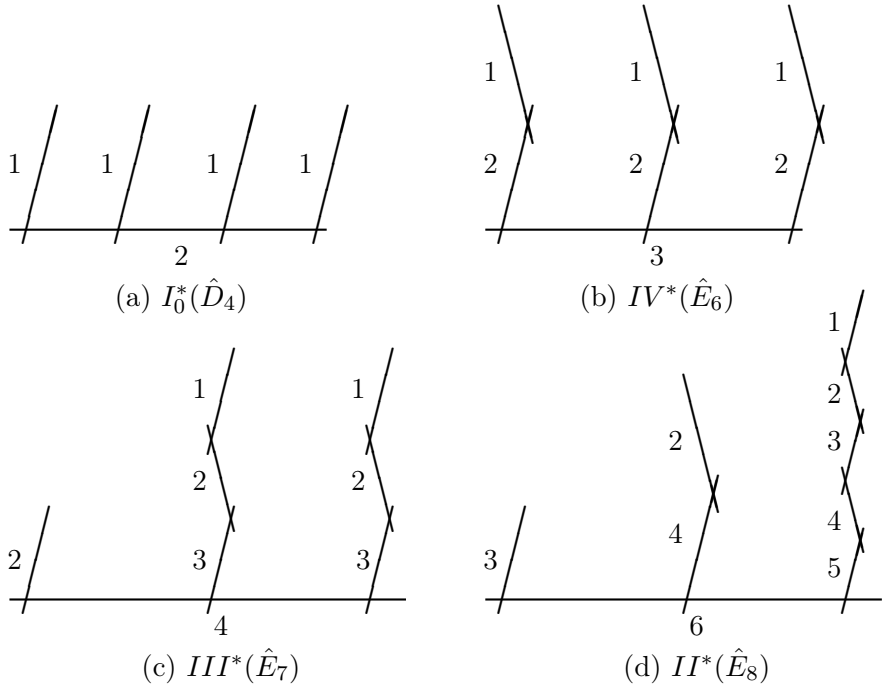
\begin{figure}
\begin{center}
\begin{picture}(200,200)(- 100,- 200)
\thicklines
% First New Structure
\put(-150, -40){\line(1, 0){120}}
\put(-85,-65){\makebox(0,0){(a) $I_0^* (\hat{D}_4)$}}
\put(-85,-50){\makebox(0,0){2}}
\put(-145, -45){\line(1, 4){13}}
\put(-148, -15){\makebox(0,0){1}}
\put(-35, -45){\line(1, 4){13}}
\put(-38, -15){\makebox(0,0){1}}
\put(-110, -45){\line(1, 4){13}}
\put(-113, -15){\makebox(0,0){1}}
\put(-70, -45){\line(1, 4){13}}
\put(-73, -15){\makebox(0,0){1}}

% Second New Structure
\put(30, -40){\line(1, 0){120}}
\put(95,-65){\makebox(0,0){(b) $IV^* (\hat{E}_6)$}}
\put(95,-50){\makebox(0,0){3}}
\put(35, -45){\line(1, 4){13}}
\put(35, 45){\line(1, -4){13}}
\put(32, -15){\makebox(0,0){2}}
\put(32, 20){\makebox(0,0){1}}
\put(90, -45){\line(1, 4){13}}
\put(90, 45){\line(1, -4){13}}
\put(87, -15){\makebox(0,0){2}}
\put(87, 20){\makebox(0,0){1}}
\put(145, -45){\line(1, 4){13}}
\put(145, 45){\line(1, -4){13}}
\put(142, -15){\makebox(0,0){2}}
\put(142, 20){\makebox(0,0){1}}

% Third New Structure
\put(-150, -180){\line(1, 0){150}}
\put(-70,-205){\makebox(0,0){(c) $III^* (\hat{E}_7)$}}
\put(-70,-190){\makebox(0,0){4}}
\put(-145, -185){\line(1, 4){10}}
%\put(-145, -125){\line(1, 4){10}}
%\put(-145, -114){\line(1, -4){10}}
\put(-148, -165){\makebox(0,0){2}}
%\put(-148, -135){\makebox(0,0){2}}
%\put(-148, -100){\makebox(0,0){3}}
\put(-75, -185){\line(1, 4){10}}
\put(-75, -125){\line(1, 4){10}}
\put(-75, -114){\line(1, -4){10}}
\put(-78, -165){\makebox(0,0){3}}
\put(-78, -135){\makebox(0,0){2}}
\put(-78, -100){\makebox(0,0){1}}
%\put(-40, -185){\line(1, 4){10}}
%\put(-40, -125){\line(1, 4){10}}
%\put(-40, -114){\line(1, -4){10}}
%\put(-43, -165){\makebox(0,0){2}}
%\put(-43, -135){\makebox(0,0){3}}
%\put(-43, -100){\makebox(0,0){1}}
\put(-15, -185){\line(1, 4){10}}
\put(-15, -125){\line(1, 4){10}}
\put(-15, -114){\line(1, -4){10}}
\put(-18, -165){\makebox(0,0){ 3}}
\put(-18, -135){\makebox(0,0){2}}
\put(-18, -100){\makebox(0,0){1}}

% Fourth New Structure
\put(30, -180){\line(1, 0){150}}
\put(110,-205){\makebox(0,0){(d) $II^* (\hat{E}_8)$}}
\put(110,-190){\makebox(0,0){6}}
\put(35, -185){\line(1, 4){10}}
\put(32, -165){\makebox(0,0){3}}
\put(105, -185){\line(1, 4){13}}
\put(105, -95){\line(1, -4){13}}
\put(102, -155){\makebox(0,0){4}}
\put(102, -120){\makebox(0,0){2}}
\put(165, -185){\line(1, 4){8}}
\put(162, -168){\makebox(0,0){5}}
\put(165, -130){\line(1, -4){8}}
\put(162, -150){\makebox(0,0){4}}
\put(165, -140){\line(1, 4){8}}
\put(162, -120){\makebox(0,0){3}}
\put(165, -85){\line(1, -4){8}}
\put(162, -102){\makebox(0,0){2}}
\put(165, -95){\line(1, 4){8}}
\put(162, -75){\makebox(0,0){1}}
\end{picture}
\end{center}
\caption[x]{\footnotesize Configurations of $-2$ curves associated
  with Kodaira-type surface singularities associated with degenerate
  elliptic fibers.  For these configurations, the number of fixed
  moduli associated with $-2$ curves is reduced by one.
The numbers given are the weightings needed to give an elliptic curve
with vanishing self-intersection.
Labels correspond to Kodaira singularity type and associated Dynkin diagram.}
\label{f:2-curves}
\end{figure}

The one remaining subtlety in this general picture is that for general
$\C^*$-bases some combinations of $-2$ curves must be treated
specially \footnote{Thanks to David Morrison for discussions on this
point}.  In particular, for
certain configurations of intersecting $-2$ curves associated with
Kodaira-type surface singularities there is a linear combination that
describes a degenerate genus one curve \cite{bhpv}.  In these cases, the extra
deformation directions associated with the $-2$ curves are not
independent and the contribution from $N_{-2}$ to
\eq{eq:hyper-counting} is reduced by 1.  The $-2$ curve configurations
of these types that appear in $\C^*$-bases are shown in
Figure~\ref{f:2-curves}.
These $-2$ curve configurations can be identified as those where an integral
linear combination of the $-2$ curves is a divisor with vanishing
self-intersection.  For this to occur, the
weighting of any $-2$ curve $C_i$ must be 1/2 the total of the weightings of
the  $-2$ curves that intersect $C_i$.  Some simple combinatorics
shows that the configurations in Figure~\ref{f:2-curves} are the only
possible geometries satisfying this condition that have a single $-2$
curve that intersects more than two others.\footnote{One other class
  of configurations satisfies this condition, the
type $I_b^*$ singularity, with two $-2$ curves
  each intersecting three $-2$ curves and connected by a single chain
  of $-2$ curves, but this cannot appear in a $\C^*$-base since the
  chain associated with each fiber must contain at least one $-1$
  curve, and in this case the chain connecting the two -2 curves with
  triple branching contains itself only $-2$ curves.}
Thus, for $\C^*$-bases that are not toric ($N \geq 3$), the formula
\eq{eq:hyper-counting} 
is replaced by
\begin{equation}
H_{\rm  neutral} = W-w_{\rm aut} +  (N -3) +
N_{-2}
-G_1  \;\;\;\;\;
(\C^*)\,,
\label{eq:hyper-counting-general}
\end{equation}
where  $w_{\rm aut} = 1 + {\rm max} (0, 1 + n_0, 1 + n_\infty)$,
$N_{-2}$ is the number of -2 curves,
and
$G_1$ is the number of $-2$ configurations of the types shown in
Figure ~\ref{f:2-curves}.

\subsection{Distribution of Hodge numbers}

\begin{figure}
\centering
\begin{picture}(200,240)(- 100,- 120)
\put(0,0){\makebox(0,0){\includegraphics[width=12cm]{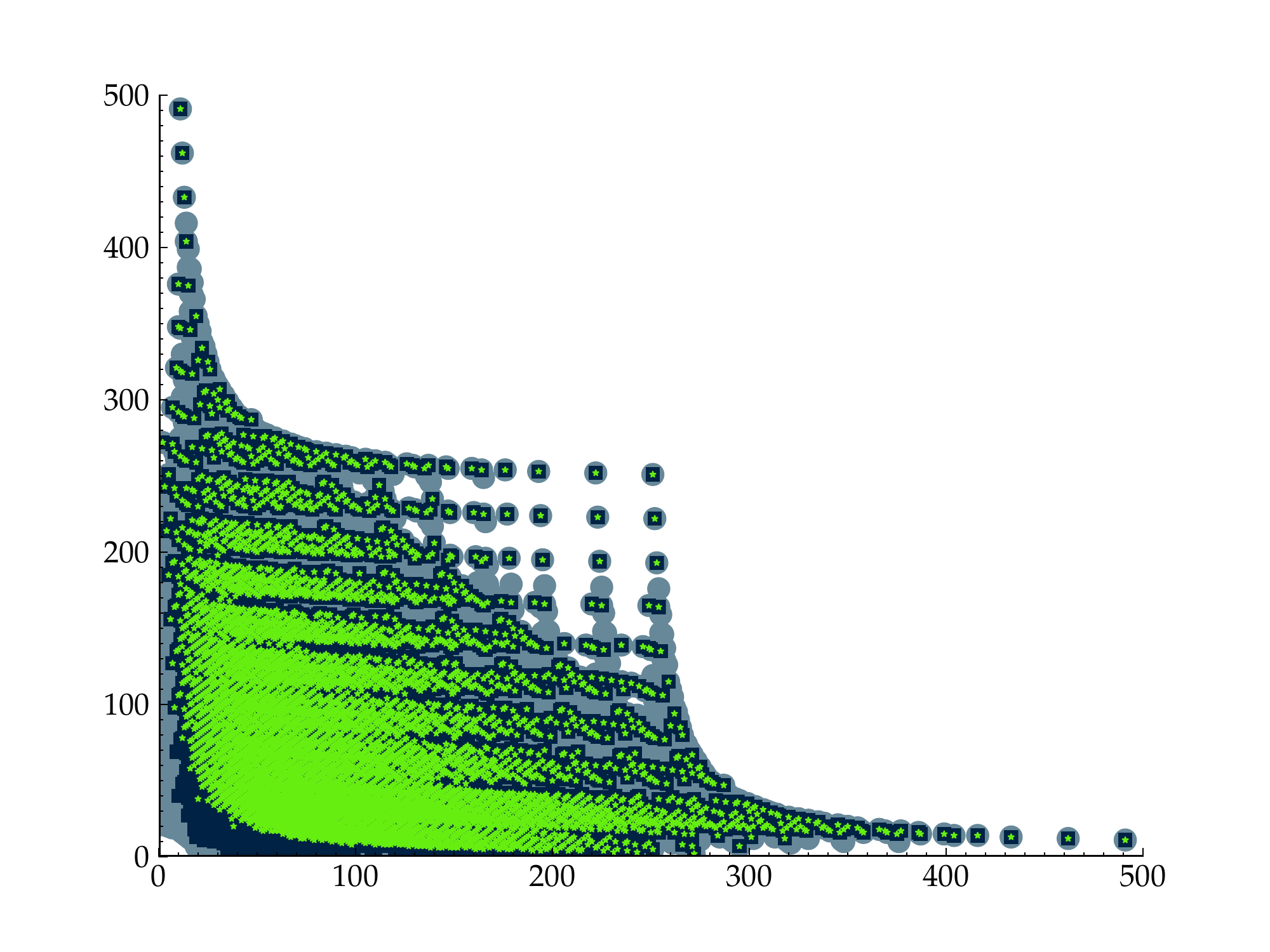}}}
\put(160,-98){\makebox(0,0){ $h^{1, 1}$}}
\put( -150,90){\makebox(0,0){ $h^{2, 1}$}}
\end{picture}
\caption[x]{\footnotesize Plot of Hodge numbers comparing $\C^*$
  (shown in dark blue, largest dots),
  Kreuzer-Skarke (slate blue, medium dots), and toric (green, smallest
  dots). 
Hodge numbers for $\C^*$-bases that are not also Hodge numbers for
toric (or NS-toric) bases are generally those with small $h^{1, 1},
h^{2, 1}$ with a small number of exceptions having larger Hodge
numbers, including 6 examples that are not found in the Kreuzer-Skarke
database.
%I wasn't able to label the axes due to some issues in my plotting
%module. I'm not sure how to fix this issue at the moment. I think the
%file may also be somewhat large.
} 
\label{f:Hodge}
\end{figure}

\begin{figure}
\begin{center}
\includegraphics[width=12cm]{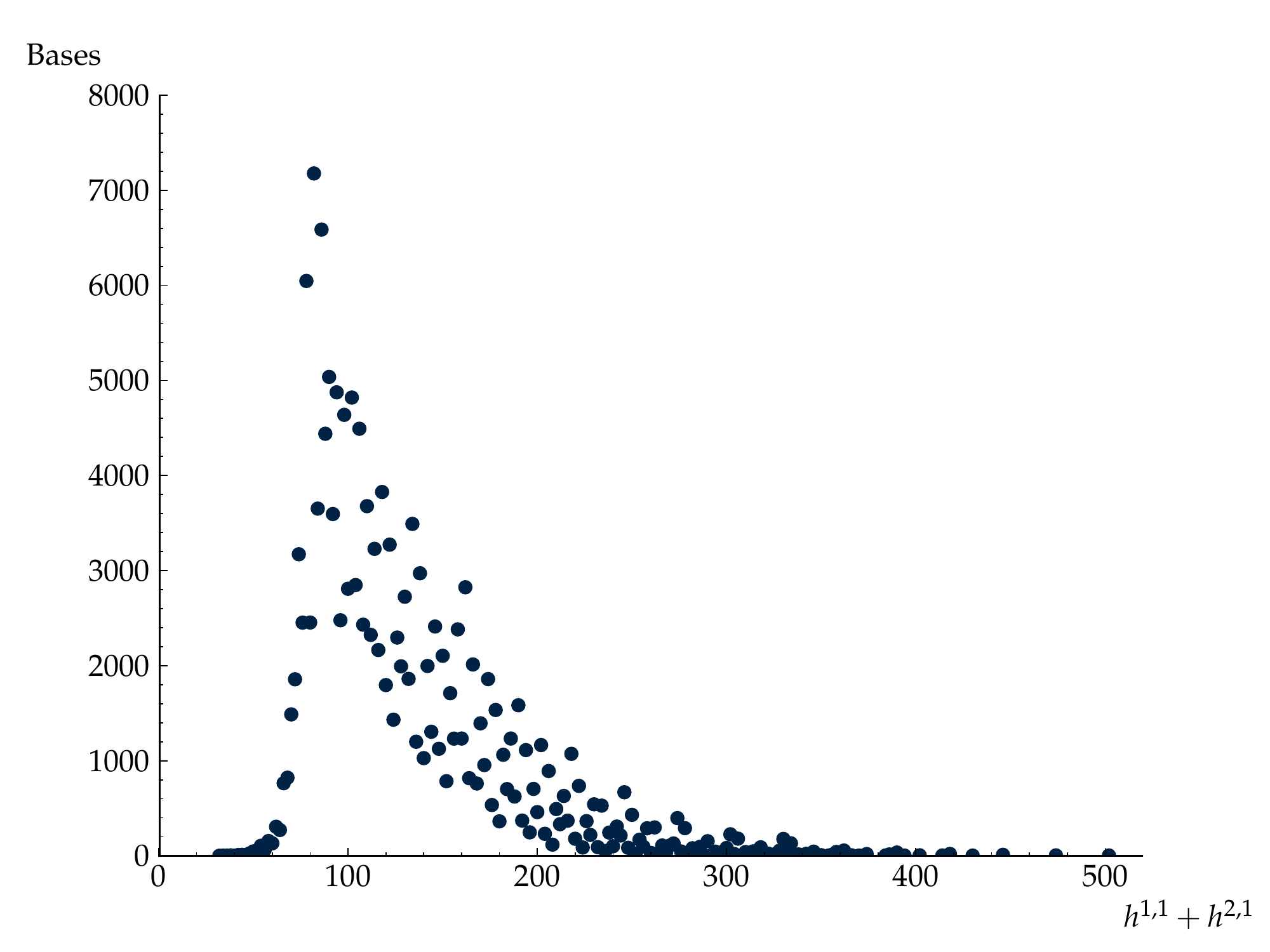}
\end{center}
\caption[x]{\footnotesize   Distribution of threefolds coming from
  generic elliptic fibrations over $\C^*$-bases as a function of the
  sum of Hodge numbers.}
\label{f:Hodge-h}
\end{figure}

We have computed the Hodge numbers for the generic threefolds over all
162,404 $\C^*$-bases, including those not strictly $\C^*$-bases coming
from blown up $\C^*$-bases with $-9, -10,$ and $-11$ curves on the
fiber chains.  We find a total of 7,868 distinct pairs of Hodge numbers
$\ho, \htt$, including 344  Hodge number
% WT reduced both numbers by 4, since 28, 10 seems also new.
combinations not found in 
\cite{WT-Hodge} from the set of generalized toric bases.  These Hodge
numbers are plotted and compared to the Kreuzer-Skarke database
\cite{Kreuzer-Skarke} and the Hodge numbers of threefolds associated
with toric bases in Figure~\ref{f:Hodge}.  The
number of models as a function of the sum $h^{1, 1} + h^{2, 1}$ is
plotted in Figure~\ref{f:Hodge-h}.
Most of the new Hodge numbers that appear for $\C^*$-bases and not for
toric bases are in the
region of small Hodge numbers far from the boundary.

Note that just as for toric bases, as discussed in \cite{toric},
the set of Calabi-Yau manifolds that can be constructed over any given
$\C^*$-base $B$ can be very large.  By tuning parameters in the
Weierstrass model over any given base, so that $f$ and $g$ vanish on
certain divisors to degrees less than $4, 6$, theories with many
different nonabelian gauge group factors can be constructed.  Each
such construction gives a different Calabi-Yau threefold after the
singularities in the elliptic fibration are resolved.  In this way,
the number of Hodge numbers associated with Calabi-Yau threefolds
fibered over any given $\C^*$-base can be quite large.  
Explicit examples of such tunings that give Hodge numbers near the
boundary of the ``shield'' for threefolds over toric bases are
described in \cite{WT-Hodge, WT-Sam}.  Similar constructions over
$\C^*$-bases would give a vast range of different Calabi-Yau threefold
constructions. 

\subsection{Redundancies from $-2$ clusters}
\label{sec:redundancies}

One striking feature of the distribution of Hodge numbers is that
there are certain Hodge number combinations that are realized by the
threefolds associated with a large number of distinct $\C^*$-bases.
As the most extreme example, there are 1,861
different  $\C^*$-bases with Hodge numbers $43, 43$.  Many of these are in
fact just different realizations of the same Calabi-Yau threefold.

\begin{figure}
\includegraphics[width=12cm]{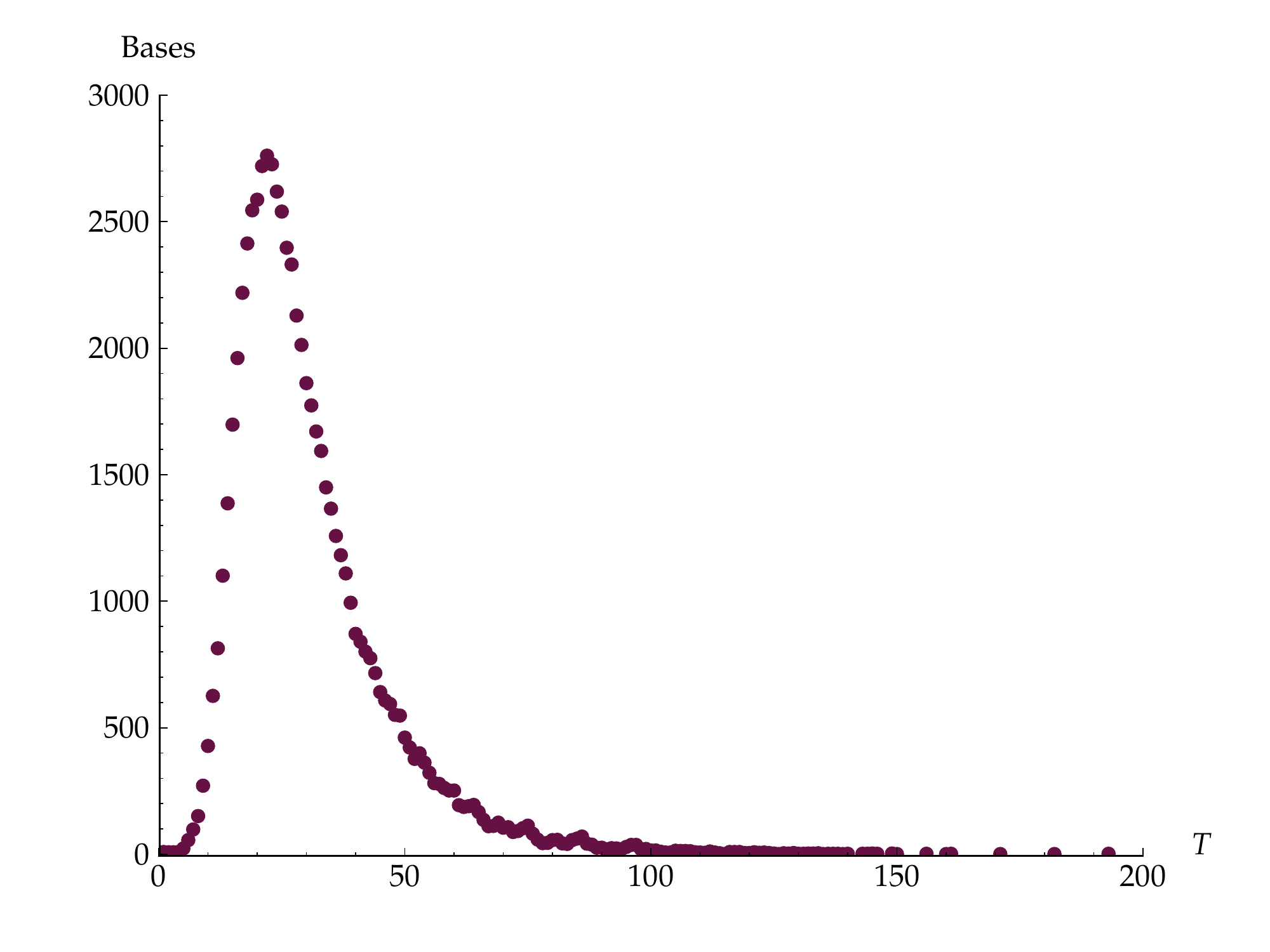}
\caption[x]{\footnotesize The number of $\C^*$-bases associated
  with different values of $T = h^{1, 1} (B) -1$, when only bases without clusters of
  $-2$ curves not carried a gauge group are considered.}
\label{f:distribution-no2}
\end{figure}

One principal source of these kinds of redundancies arises from the
appearance of clusters of $-2$ curves in the base that do not carry a
gauge group.  As discussed previously,  such $-2$ curves indicate
that a modulus of the geometry has been tuned.  In general, we expect
that any base containing a cluster of $-2$ curves is just be a special
limit of another base without such clusters.  Thus, we can consider
the subset of $\C^*$-bases that do not contain any clusters of only
$-2$ curves.  This reduces the number of bases to 68,798.
% WT: reduced by 3 since old c.3 had -2 cluster.
  A graph of
the distribution of the numbers of bases without $-2$ clusters as a
function of $T$ is given in Figure~\ref{f:distribution-no2}.

The removal of $-2$ clusters  removes a great deal of redundancy in
the list of threefolds.
In particular, in the reduced set of bases the extreme jump in the
distribution at $T = 25$ goes away.  This is associated with the
removal of a large number of threefolds with  Hodge numbers
$(43, 43)$.  Looking at the detailed data shows that many of these
$(43, 43)$
models have a closely related structure.
There are 1,575 $\C^*$-bases that have the following
features:
\begin{itemize}
\item Two $(-12)$ clusters and a gauge algebra $\ge_8
  \oplus\ge_8$
\item $T = 25$
\item $\ho = 43, \htt = 43$ \,.
\end{itemize}
In general, these bases are characterized by $\ge_8$ factors on the
sections ($n_{0, \infty} = -12$), and a set of chains containing
various combinations of $-2$ curves.  Indeed, it is clear that there
are many ways to construct such $\C^*$-bases, by starting with a
given $\F_m$ and blowing up only points at $w = 0, \infty$.  Such
$\C^*$-bases will always have chains of the form $(-1, -2, -2,
\ldots, -2, -1)$.  If precisely 24 points are blown up, giving the
necessary factors on the sections, then any combination of chain
lengths satisfying 
\begin{equation}
\sum_{i = 1}^{N}  (k_i-1) = 24 = T-1
% \label{eq:}
\end{equation}
will have the desired properties.  The number of such bases is just
the number of ways of partitioning the 24 blow-ups into a sum of
integers, $p (24) = 1,575$, precisely the number of bases found with
the above features.  Of these partitionings, 13 have toric
descriptions (partitions into one or two integers).  All 1575
$\C^*$-bases in this set can be understood as limit points of the
same geometry, and are associated with the same smooth Calabi-Yau
threefold.  This Calabi-Yau threefold, which has been encountered
previously in the literature (see {\it e.g.} \cite{Candelas-pr}),
seems to have a particularly high degree of symmetry and may be
interesting for other reasons.

Although many of the bases with clusters of $-2$ curves have the same
Hodge numbers as bases without such clusters, this is not universally
true.  There are also 3,788 $\C^*$bases that have
% WT reduced by 1 because of old c.3
$-2$ clusters that have no corresponding $\C^*$-base without such
clusters.  An example is given by base $B$ with 
\begin{eqnarray}
N = 4, &  & \; n_0 = -2, n_\infty = -4\nonumber \\
{\rm chain}\ 1: &  &  (-2, -1, -4, -1, -3, -1)\nonumber \\
{\rm chains}\  2-4: &  &  (-1, -1) \nonumber
\end{eqnarray}
These bases may be
limits of other base surfaces that are fine as complex surfaces but do
not have a description as a $\C^*$-surface.

While as discussed above, in general bases with
$-2$ clusters correspond to limits of bases without such clusters, and
amounts to a redundancy for {\em generic} elliptic fibrations,
keeping track of this information is often useful.
In particular,  Weierstrass models over a base with $-2$ clusters can
be tuned to realize smooth resolved Calabi-Yau threefolds that cannot
be realized by tuned Weierstrass models over the more generic bases
without the $-2$ clusters.  Thus, bases with $-2$ clusters must be
considered separately in any systematic or complete analysis of a set
of Calabi-Yau threefolds; examples of this arise in
\cite{WT-Sam}.  
In terms of the physical F-theory models that can be constructed from
these bases, the different $-2$ configurations and the distinct
configurations of tuned gauge groups that can be realized over them
correspond to distinct classes of 6D supergravity theories with
distinct gauge group, matter, and dyonic string lattice structure.
Thus, in a full consideration of 6D supergravity theories, it would be
necessary to include all of the distinct base choices with Hodge
numbers (43, 43) as initial points for tuned models with different
sets of Hodge numbers.

\subsection{Calabi-Yau threefolds with new Hodge numbers}
\label{sec:Hodge}

Of the roughly 160,000
$\C^*$-bases, we have found precisely 6  that give rise to generic
elliptic fibrations with Hodge numbers that are not found in the
Kreuzer-Skarke database.  The simplest such base has
Hodge numbers
$\ho = 56, \htt =  2$, and the structure

\vspace*{-0.05in}
\begin{eqnarray}
N = 3, &  & 
n_0 = -5, n_\infty = -6, \label{eq:new-1}
\\
{\rm chain}\ 1: &  & (-1, -3, -1, -3, -1) \nonumber\\
{\rm chain}\ 2: &  & (-1, -3, -2, -1, -5, -1, -3, -1)\nonumber\\
{\rm chain}\ 3: &  & (-1, -3, -2, -2, -1, -6, -1, -3, -1) \nonumber
\end{eqnarray}

The other 5 bases that give new Calabi-Yau threefolds are listed in
Appendix \ref{sec:new-threefolds}.

\subsection{Calabi-Yau threefolds with nontrivial Mordell-Weil rank}
\label{sec:Mordell-Weil}

There are 13 $\C^*$-bases in which the rank of the Mordell-Weil group
is nonzero.  This is determined, as described above, by using the
monomial count and the anomaly equation to independently determine
$H_{\rm neutral}$ and $H_{\rm neutral} -V_{\rm abelian}$
for each of the $\C^*$-bases.  The bases where these two quantities
differ are those
for which the Mordell-Weil rank is nonzero.
The elliptically fibered Calabi-Yau threefolds over these
bases have multiple linearly independent sections in a group of rank
$r$.  This gives rise to $r$ abelian $U(1)$ gauge fields in the
corresponding 6D supergravity theory.  

For these 13 bases, therefore, the Mordell-Weil rank is forced to be
nontrivial even for a completely generic elliptic fibration.  This is
different from the situation for toric bases, where a complete
analysis of all toric bases using the anomaly condition
\eq{eq:anomaly} confirmed that there are no toric bases over which the
generic elliptic fibration has a nontrivial Mordell-Weil rank.

An example of a base with nonzero Mordell-Weil rank is given by
the following base with
Hodge numbers
$\ho = 25, \htt = 13$:
\vspace*{-0.05in}
\begin{eqnarray}
N = 3, &  & 
n_0 = -1, n_\infty = -2,  \label{eq:mw-example}
\\
{\rm chain}\ 1: &  &  (-2, -1, -2) \nonumber\\
{\rm chain}\ 2: &  & (-4, -1, -2, -2, -2)\nonumber\\
{\rm chain}\ 3: &  & (-4, -1, -2, -2, -2) \nonumber
\end{eqnarray}
The Hodge number $\ho$ can be computed from this base using
\eq{eq:h11}, and $T = 11$ from \eq{eq:t-equation}, where ${\cal G}
=\gso(8)\oplus\gso (8)$ from the two $-4$ curves, so $\ho (X) = T + 2
+ 8 + r = 21 + r$, where $r$ is the rank of the Mordell-Weil group.
Using the method of \S\ref{sec:neutral}, the number of Weierstrass
monomials can be computed to be $W = 7$.  The dimension of the
automorphism group is the generic $w_{\rm aut} = 1$, and $N_{-2} = 9$
is the number of $-2$ curves, with 
$G_1 = 1$ as described at the
end of \S\ref{sec:neutral} since the $-2$ curves connected to
$D_\infty$ have the $IV^* (\hat{E}_6)$ form from Figure~\ref{f:2-curves}.
It follows then from
\eq{eq:hyper-counting-general} that $H_{\rm neutral} = 14$, so $\htt =
13$.  Comparing with \eq{eq:anomaly}, we have 
\begin{equation}
V= H_{\rm neutral} + 29T-273 = 60  = 56 + r\,.
% \label{eq:}
\end{equation}
Thus, this base has a generic elliptic fibration with Mordell-Weil
rank $r = 4$, and $\ho = 25$.
The 13 examples of bases with Mordell-Weil groups having nonzero
rank are listed in Appendix~\ref{sec:appendix-abelian}

The appearance of such bases, while perhaps surprising, is not
completely unprecedented.  It is known that the Shoen manifold, a
class of elliptically fibered Calabi-Yau threefold constructed from a
fiber product of rational elliptic surfaces, generically has
Mordell-Weil rank 9 \cite{Shoen} (see {\it e.g.} \cite{bopr} for a
physics application of this).  In fact, all the $\C^*$-bases we have
identified that give Calabi-Yau threefolds with enhanced Mordell-Weil
rank can be related to special limits and blow-ups of the Shoen
manifold\footnote{Thanks to David Morrison for discussions on this
  point.}; this connection will be described in further detail
elsewhere \cite{mpt}.

For  the bases with nonzero Mordell-Weil rank, an explicit description
of the Weierstrass model can be used to make the extra sections
manifest.  
As an example, consider the base
\begin{eqnarray}
N =  4, &  & 
n_0 = -6, n_\infty = - 6,  \label{eq:mw-example-2}
\\
{\rm chain}\ 1: &  &   (-1, -3, -1, -3, -1) \nonumber\\
{\rm chain}\ 2: &  &   (-1, -3, -1, -3, -1) \nonumber\\
{\rm chain}\ 3: &  &   (-1, -3, -1, -3, -1) \nonumber\\
{\rm chain}\ 4: &  &   (-1, -3, -1, -3, -1) \nonumber
\end{eqnarray}
This base has $T = 17$, $h^{1, 1}= 51, h^{2, 1}= 3$, and Mordell-Weil
rank $r = 4$.  An explicit computation of the monomials in the
Weierstrass model gives (placing the extra two fibers at $z = 1, 2$)
\begin{eqnarray}
f & = & Aw^4z^2(z-1)^2(z-2)^2 \label{eq:11-Weierstrass}\\
g & = & Bw^4z^6(z-1)^6(z-2)^6 + Cw^6z^3(z-1)^3(z-2)^3  + Dw^8 \nonumber   \,,
\end{eqnarray}
where $A, B, C, D$ are free complex constants ($W = 4$).  A nontrivial
section can be associated with a factorization of
the Weierstrass equation $y^2 = x^3 + fx + g$ into the form
\cite{Morrison-Park}
\begin{equation}
(y-\alpha) (y + \alpha) = (x-\lambda) (x^2 + \lambda x-\mu)
% \label{eq:}
\end{equation}
where
\begin{eqnarray}
f & = &  -\mu -\lambda^2\\
g & = &  \lambda \mu + \alpha^2 \,.
\end{eqnarray}
The Weierstrass coefficients from (\ref{eq:11-Weierstrass}) can take
this form if
\begin{eqnarray}
\lambda & = & aw^2z(z-1)(z-2)\\
\mu & = & bw^4z^2(z-1)^2(z-2)^2\\
\alpha & = & cw^2z^3(z-1)^3(z-2)^3 + dw^4
\end{eqnarray}
where $a, b, c, d$ satisfy
\begin{eqnarray}
A & = & -a^2 -b\\
B & = & c^2\\
C & = & ab + 2cd\\
D & = & d^2\,.
\end{eqnarray}
For given generic values of $A$-$D$ there are 12 solutions for
$a$-$d$.  This can be seen by noting that the equations for $B, D$
each have two solutions for $c, d$, and the other two equations
combine to form a cubic for $a$, which has 3 independent solutions.
These twelve solutions represent 4 independent generators of the
Mordell-Weil group.  Solutions with $\alpha \rightarrow -\alpha$
correspond to sections $s, -s$ that add to 0.  The three solutions for
the cubic also contain one linear dependence in the space of sections,
so that the total number of independent sections is 4, matching the
computed Mordell-Weil rank.
A similar computation can be carried out for the other bases
with enhanced Mordell-Weil rank, though the details are more
complicated and computing the number of independent sections can be
more difficult in other cases.  A more detailed analysis of these
models with enhanced Mordell-Weil rank is left to future work.

\section{Conclusions}
\label{sec:conclusions}

In this paper we have initiated a systematic study of a class of
geometries for the bases of elliptically fibered Calabi-Yau threefolds
that goes beyond the framework of toric geometry widely used in
previous work.
We have systematically constructed all smooth surfaces that admit a
single $\C^*$ action and can arise as bases of a Calabi-Yau threefold,
and we have analyzed the properties of these geometries.  The 162,404
bases we have explicitly constructed include all $\C^*$-bases and also
a more general class built from $\C^*$-bases containing $-9, -10,$ and
$-11$ curves on which points must be blown up to form $-12$ curves in
bases that do not have a $\C^*$ action.

The bases we have considered can be used for compactification of
F-theory to six dimensions.  We have found that the physical
properties of the resulting six-dimensional supergravity theories are
similar in nature and in distribution
to compactifications on toric bases that were
studied earlier.  In particular, the $\C^*$-bases with relatively
large numbers $T$ of tensor multiplets give theories with gauge
algebras that are dominated by summands of the form $\ge_8 \oplus\gf_4
\oplus 2 (\gg_2 \oplus\gsu (2))$, from the same types of chains that
give this structure in the toric case.  The largest value of $T$ for
theories with $\C^*$-bases that are not actually toric is $171$, lower
than the largest known value of $T = 193$ that occurs for a toric base
(in both cases, these are examples of bases containing $-11$ curves
that must be blown up to give bases outside the $\C^*$/toric
framework).  The overall ``shield'' structure of the set of Hodge
numbers computed by Kreuzer and Skarke based on toric constructions,
the boundary of this region explained in \cite{toric}, and geometric
patterns identified in \cite{toric, Candelas-cs} are essentially
unchanged with the addition of the large number of more general
non-toric $\C^*$-base configurations.
The absence of branching or loop structures in the
$\C^*$-bases that make possible higher values of $T$ supports the
conclusion that $T \leq 193$ is an absolute bound across all bases, as
was argued heuristically in \cite{toric}.

One interesting result of this analysis is that the number of
additional bases added by extending the toric construction to include
the more general class of bases admitting a $\C^*$ action does not
produce a wild increase in the number of possible bases.  As more
points are blown up, the set of geometries is generally controlled by
linear structures along the fiber, and the possibility of branching
does not lead to a combinatorial explosion in intersecting divisor
structures.  This suggests that a systematic classification of all
smooth bases for elliptically fibered Calabi-Yau threefolds, even
without a single $\C^*$ action, may be computationally tractable.
Such a classification would be quite challenging, since the
intersection structure can become quite complicated -- particularly
for $T \geq 9$ where there  can be an infinite number of distinct $-1$
curves on the bases, such as occurs for dP$_9$.  Nonetheless, by using the
method of non-Higgsable clusters to characterize possible geometries
it may be possible to get a handle on this problem.  We leave this as
a challenge for future work.

We have analyzed the Hodge structure of the smooth Calabi-Yau
threefolds associated with generic elliptic fibrations over all the
$\C^*$-bases.  While the Hodge numbers are all within the boundaries
defined by toric bases and the Kreuzer-Skarke ``shield'' shape, we
have identified a number of Calabi-Yau threefolds with novel
properties.  There are 6 new threefolds that have Hodge numbers
that we believe have not been previously identified, as well as 13
threefolds in which the Mordell-Weil rank of the elliptic fibration is
nontrivial, corresponding to non-Higgsable $U(1)$ factors in the
corresponding 6D supergravity theory.

The basic approach to constructing a more general class of base
manifolds for elliptic fibrations and F-theory using spaces with
reduced numbers of $\C^*$ actions should be possible for base
threefolds and fourfolds
as well, giving rise to geometries for compactification of F-theory
and M-theory to dimensions from 5 to 2.  The analysis of these
constructions becomes significantly more complicated in lower
dimensions, however.  In particular, for F-theory compactifications to
four dimensions on elliptically-fibered Calabi-Yau fourfolds with
threefold bases, the analysis of bases with a $\C^*\times\C^*$ or
$\C^*$ action would be much more difficult than for the class of
$\C^*$-surfaces considered here.
One difficulty in analyzing such constructions is the absence of a
clear analogue to the anomaly condition \eq{eq:anomaly} for 6D
theories.  The absence of such a condition makes it harder to compute
the Mordell-Weil group in a general 4D case, and hence more difficult to
precisely identify the Hodge numbers of the elliptically-fibered
fourfold.  
While some progress was made in \cite{Grimm-Taylor} in identifying 4D
parallels to the 6D anomaly condition and related structures, more
work is needed to have a systematic approach to analyzing F-theory
base spaces with reduced toric symmetry in the 4D case.

\vspace*{0.1in}

{\bf Acknowledgements}: We would like to thank Lara Anderson, Philip
Candelas, Mboyo Esole, Ilarion Melnikov, David Morrison, Daniel Park,
Tony Pantev, and Yinan Wang for helpful discussions.  This research was supported
by the DOE under contract \#DE-FC02-94ER40818, and was also supported
in part by the National Science Foundation under Grant
No. PHY-1066293.  WT would like to thank the Aspen Center for Physics
for hospitality during part of this work.

\appendix

\section{Rules for connecting clusters in $\C^*$-bases}
\label{sec:rules}

\begin{table}
\begin{center}
\begin{tabular}{| r | 
r |l |
}
\hline
Index ($n$) & NHC &  pairs \{triplets [quadruplets]\}
\\
\hline
\hline
1 & ($-\dot{12}$) & 12  \\
2 & ($-\dot{8}$) & 9, 12, 13   \\
3 & ($-\dot{7}$) & 9, 12, 13  \\
4 & ($-\dot{6}$)  & 7, 9, 11, 12, 13   \\
5 & ($-\dot{5}$)  & 7, 8, 9, 10, 11, 12, 13 \\
6 & ($-\dot{4}$)  & 6, 7, 8, 9 \{12\}, 10, 11, 12 \{12, 13\}, 13, 14 \\
7 & ($-\dot{3}$)  & 7 \{7, 9, 12\}, 8, 9 \{9, 12, 13\}, 10, 11, 12 \{12, 13\}, 13 \{13\}, 14 \\
8 & ($-\dot{3}, -2$) & 8, 9 \{9, 12\}, 10, 11, 12 \{12, 13\}, 13, 14 \\
9 & ($-3, -\dot{2}$) & 9 \{9, 10, 12 [12], 13\}, 10 \{12\}, 11, 12 \{12 [12, 13], 13\}, 13 \{13\}, 14 \\
%& & $\,\,\,\,\,\,\,$ 12 \{12 [12, 13], 13\}, 13 \{13\}, 14   \\
10 & ($-\dot{3}, -2, -2$) & 10, 11, 12 \{12, 13\}, 13, 14 \\
11 & ($-3, -\dot{2}, -2$) &  12 \{12, 13\}, 13 \{13\} \\
12 & ($-3, -2, -\dot{2}$) & 12 \{12 [12, 13], 13 [13]\}, 13 \{13\}, 14  \\
13 & ($-\dot{2},-3, -2$) & 13 \{13\}, 14  \\
14& ($-2, -\dot{3}, -2$) & \\

\hline
\end{tabular}
\end{center}
\caption[x]{\footnotesize Table of all ways in which connections
between non-Higgsable clusters by (-1)-curves are allowed to appear in
a $\C^*$-surface that can be used for an F-theory base. The dots
specify the curves within each NHC that intersect with the
(-1)-curve. The last column provides a complete list of the allowed
pairs, triplets and quadruplets of NHCs that can be intersected by a
single (-1)-curve in a valid F-theory base. No more than 4 NHCs can be
connected by a single (-1) curve.
Note also that any of the single NHCs or set of several NHCs
that can be intersected by a single (-1)-curve can also be connected
by the same (-1)-curve to sets of -2 curves, which do not carry a
nonabelian gauge group. 
This table contains the subset of the data from an analogous table in
\cite{clusters} that is relevant for $\C^*$-bases.
Note that combinations are listed only in increasing numerical order,
so that a given combination only appears on the row associated with the
lowest numbered NHC in the connected set.
}
\label{t:NHCs}
\end{table}

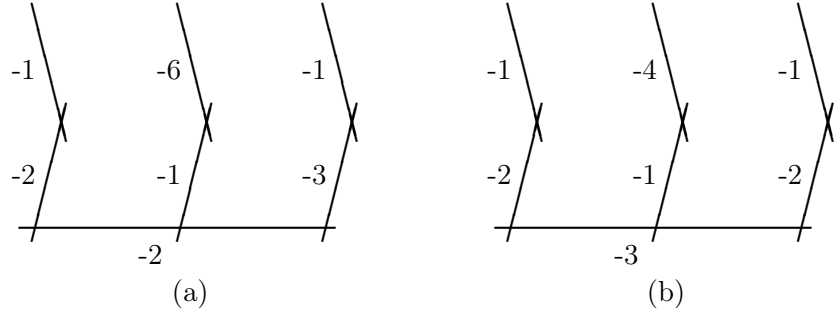
\begin{figure}
\begin{center}
\begin{picture}(200,100)(- 100,- 50)
\thicklines
% First New Structure
\put(-150, -40){\line(1, 0){120}}
\put(-85,-65){\makebox(0,0){(a)}}
\put(-100,-50){\makebox(0,0){-2}}
\put(-145, -45){\line(1, 4){13}}
\put(-145, 45){\line(1, -4){13}}
\put(-148, -20){\makebox(0,0){-2}}
\put(-148, 20){\makebox(0,0){-1}}
\put(-90, -45){\line(1, 4){13}}
\put(-90, 45){\line(1, -4){13}}
\put(-93, -20){\makebox(0,0){-1}}
\put(-93, 20){\makebox(0,0){-6}}
\put(-35, -45){\line(1, 4){13}}
\put(-35, 45){\line(1, -4){13}}
\put(-38, -20){\makebox(0,0){-3}}
\put(-38, 20){\makebox(0,0){-1}}

% Second New Structure
\put(30, -40){\line(1, 0){120}}
\put(95,-65){\makebox(0,0){(b)}}
\put(80,-50){\makebox(0,0){-3}}
\put(35, -45){\line(1, 4){13}}
\put(35, 45){\line(1, -4){13}}
\put(32, -20){\makebox(0,0){-2}}
\put(32, 20){\makebox(0,0){-1}}
\put(90, -45){\line(1, 4){13}}
\put(90, 45){\line(1, -4){13}}
\put(87, -20){\makebox(0,0){-1}}
\put(87, 20){\makebox(0,0){-4}}
\put(145, -45){\line(1, 4){13}}
\put(145, 45){\line(1, -4){13}}
\put(142, -20){\makebox(0,0){-2}}
\put(142, 20){\makebox(0,0){-1}}

% Third New Structure
%\put(-75, -180){\line(1, 0){150}}
%\put(5,-205){\makebox(0,0){(c)}}
%\put(-10,-190){\makebox(0,0){-1}}
%\put(-70, -185){\line(1, 4){10}}
%\put(-70, -125){\line(1, 4){10}}
%\put(-70, -114){\line(1, -4){10}}
%\put(-73, -165){\makebox(0,0){-2}}
%\put(-73, -135){\makebox(0,0){-2}}
%\put(-73, -100){\makebox(0,0){-3}}
%\put(-35, -185){\line(1, 4){10}}
%\put(-35, -125){\line(1, 4){10}}
%\put(-35,-114){\line(1, -4){10}}
%\put(-38, -165){\makebox(0,0){-2}}
%\put(-38, -135){\makebox(0,0){-2}}
%\put(-38, -100){\makebox(0,0){-3}}
%\put(0, -185){\line(1, 4){10}}
%\put(0, -125){\line(1, 4){10}}
%\put(0, -114){\line(1, -4){10}}
%\put(-3, -165){\makebox(0,0){-2}}
%\put(-3, -135){\makebox(0,0){-3}}
%\put(-3, -100){\makebox(0,0){-1}}
%\put(35, -185){\line(1, 4){10}}
%\put(35, -125){\line(1, 4){10}}
%\put(35, -114){\line(1, -4){10}}
%\put(32, -165){\makebox(0,0){-2}}
%\put(32, -135){\makebox(0,0){-3}}
%\put(32, -100){\makebox(0,0){-1}}
%\put(60, -185){\line(1, 4){10}}
%\put(60, -125){\line(1, 4){10}}
%\put(60, -114){\line(1, -4){10}}
%\put(57, -165){\makebox(0,0){-1}}
%\put(57, -135){\makebox(0,0){-12}}
%\put(57, -100){\makebox(0,0){-1}}
\end{picture}
\end{center}

\caption[x]{\footnotesize Examples of how clusters connected by $-1$
  curves can arise at the branching points of the configuration of
divisors in $\C^*$-bases.}
\label{f:branching-examples}
\end{figure}

All configurations of non-Higgsable clusters (NHC's) that
can be connected by a single $-1$ curve in a $\C^*$-base are listed in
Table~\ref{t:NHCs}.  Some examples of how some of these configurations
can arise in $\C^*$-bases are given in Figure~\ref{f:branching-examples}.

There are a few additional constraints not analyzed in
\cite{clusters}, that arise from conditions on branching over a $-n$
curve with $n > 1$; these additional constraints arise for similar
  reasons to those underlying the constraints on branching over a $-1$ curve.
The additional constraints that must be imposed on such branchings are
as follows:

\begin{itemize}
\item[$\bullet$]{\bf $-3$ curves:}

A $-3$ curve that is connected to a $-2$ curve can be connected to at
most one chain that begins $-1, -5, \ldots$.

\item[$\bullet$]{\bf $-5$ curves:}

A $-5$ curve can connect to at most two chains that begin
$-1, -3, -2, \ldots$.

\item[$\bullet$]{\bf $-7$ curves:}

A $-7$ curve can connect to at most five chains that begin
$-1, -2, -3, \ldots$.

\item[$\bullet$]{\bf $-8$ curves:}

A $-8$ curve can connect to at most four chains that begin
$-1, -2, -3, \ldots$.

\item[$\bullet$]{\bf $-12$ curves:}

A $-12$ curve can connect to at most eight chains that begin
$-1, -2, -2, -3, \ldots$.
\end{itemize}

As an example of how these constraints arise, consider the Zariski
decomposition of $-4K, -6K$ for a sequence of rational curves $A, B,
C, D$ of self-intersections $-5, -1, -3, -2$ with mutual intersections
$A \cdot B = B \cdot C = C \cdot D$.  From the
relations $-K \cdot A = -3, -K \cdot C = -1, \ldots$ we have
\begin{eqnarray}
-4K & = &  3A + B + 2C + D + X\\
-6K & = &  4A + B + 3C + 2D +  Y
\end{eqnarray}
where $X, Y$ are residual effective divisors that have nonnegative
intersection with each of the curves $A$-$D$.  This Zariski
decomposition is compatible with the $(3, 4)$ vanishing on the $-5$
curve, etc., and implies that $(f, g)$ both vanish to order one on the
curve $B$.
Now, consider a configuration where the curve $A$ of self-intersection
$-5$ is attached to three sequences $(B, C, D)$, $(B', C', D')$,
$(B'', C'', D'')$ of curves of self-intersections $-1, -3, -2$.  Then
both
$-4K$
and $-6K$ would have to contain all of $B, B', B''$ as components.
Since, however, $(4A + B + B' + B'') \cdot A = -17$, while $-6K \cdot
A = -18$, the Zariski decomposition of $-6K$ must contain at least
$5A$, so that $g$ must vanish to order 5 on $A$.  This would mean,
however, that $(f, g)$ would vanish to orders $(4, 6)$ at the point $A
\cap B$, which we do not allow.  This demonstrates the condition
stated above for branching on a $-5$ curve.  Similar arguments can be
used to show the other conditions.

The conditions on $-3$ and $-5$ curve branching rule out some
additional candidates for $\C^{*}$-bases that are otherwise apparently
acceptable.\footnote{These additional rules were missed in an earlier
  version of this paper, which led to the appearance of four
  additional candidate $\C^{*}$-bases.  Remarkably, these four all
  were in the class of bases with generic nonzero rank for the
  Mordell-Weil group.  Thanks to D.\ Morrison,
D.\ Park, and Y.\ Wang for discussions on related points.}

\section{Bases giving Calabi-Yau threefolds with new Hodge numbers}
\label{sec:new-threefolds}

In addition to the example (\ref{eq:new-1}) given in Section
\ref{sec:Hodge}, there are 5 other $\C^*$-bases over which the generic
elliptic fibration has Hodge numbers not found in the Kreuzer-Skarke database:

\noindent
Hodge numbers
$\ho = 61, \htt = 1$:
\vspace*{-0.05in}
\begin{eqnarray}
N = 3, &  & 
n_0 = -5, n_\infty = -6, 
\\
{\rm chain}\ 1: &  & (-1, -3, -1, -3, -1) \nonumber\\
{\rm chain}\ 2: &  & (-1, -3, -2, -2, -1, -6, -1, -3, -1)\nonumber\\
{\rm chain}\ 3: &  & (-1, -3, -2, -2, -1, -6, -1, -3, -1) \nonumber
\end{eqnarray}

\noindent
Hodge numbers
$\ho = 62, \htt = 2$:
\vspace*{-0.05in}
\begin{eqnarray}
N = 3, &  & 
n_0 = -4, n_\infty = -8, 
\\
{\rm chain}\ 1: &  & (-1, -4, -1, -2, -3, -2, -1) \nonumber\\
{\rm chain}\ 2: &  & (-1, -4, -1, -2, -3, -2, -1) \nonumber\\
{\rm chain}\ 3: &  & (-1, -4, -1, -2, -3, -2, -1) \nonumber
\end{eqnarray}

\noindent
Hodge numbers
$\ho = 110, \htt = 2$:
\vspace*{-0.05in}
\begin{eqnarray}
N = 3, &  & \;\;\;\;\; n_0 = -6, n_\infty = -12,\\
{\rm chain}\ 1: &  & (-1, -3, -1, -3, -2, -2, -1) \nonumber\\
{\rm chain}\ 2: &  & (-1, -3, -1, -3, -2, -2, -1) \nonumber\\
{\rm chain}\ 3: &  & (-1, -3, -1, -6, -1, -3, -1, -6, -1, -3, -1, -5,
-1, -3, -2, -2, \nonumber\\
& & \; \; -1, -12,
 -1, -2, -2, -3, -1, -5, -1, -3, -2, -2, -1) \nonumber
\end{eqnarray}

\noindent
Hodge numbers
$\ho =  241, \htt =  31$:
\vspace*{-0.05in}
\begin{eqnarray}
N = 3, &  & \;\;\;\;\; n_0 = -1, n_\infty = -5, \\
{\rm chain}\ 1: &  & (-1, -1) \nonumber\\
{\rm chain}\ 2: &  & (-2, -1, -3, -1)\nonumber\\
 {\rm chain}\ 3: &  & 
%(-12,  -1,  -2,  -2,  -3,  -1,  -5,  -1,  -3,  -2,  -2,  -1, -11, -1, \-2,  -2,  -3,  -1,\nonumber\\& & \; \;
% -5,  -1,  -3,  -2,  -2,  -1, -12,  -1, -2,  -2,  -3,  -1,  -5,  -1,  -3,  -2,  -2,  -1,\nonumber\\& & \; \;
% -12,  -1,  -2, -2,  -3,  -1,  -5,  -1,  -3,  -2,  -2,  -1, -12,  -1,  -2,  -2, -3,  -1, \nonumber\\& & \; \;
% -5,  -1,  -3,  -2,  -2,  -1, -12,  -1,  -2,  -2,  -3, -1,  -5,  -1,  -3,  -2,  -2,  -1,\nonumber\\& & \; \;
%-12,  -1,  -2,  -2,  -3,  -1, -5,  -1,  -3,  -2,  -2,  -1, 
-12//-11//-12//-12//-12//-12//-12//
-11,  -1,  -2,  -2,  -3,  -1) \nonumber
\end{eqnarray}

where, as in \cite{WT-Hodge}, the notation $X//Y$ denotes the sequence
$X,  -1,  -2,  -2,  -3,  -1, -5,  -1, \\
 -3,  -2,  -2,  -1,  Y$.
\vspace*{0.1in}

\noindent
Hodge numbers
$\ho =  252, \htt =  30$:
\vspace*{-0.05in}
\begin{eqnarray}
N = 3, &  & \;\;\;\;\; n_0 = -1, n_\infty = -6, \\
{\rm chain}\ 1: &  & (-1, -1) \nonumber\\
{\rm chain}\ 2: &  & (-2, -1, -3, -1)\nonumber\\
{\rm chain}\ 3: &  & 
%(-12, -1, -2, -2, -3, -1, -5, -1, -3, -2, -2, -1, -11, -1, -2, -2, -3,
-1,
% -5, -1, -3, -2, -2, -1, -12, -1, -2, -2, -3, -1, -5, -1, -3, -2, -2, -1,\nonumber\\& & \; \;
% -12, -1, -2, -2, -3, -1, -5, -1, -3, -2, -2, -1, -12, -1, -2, -2, -3, -1,\nonumber\\& & \; \;
% -5, -1, -3, -2, -2, -1, -12, -1, -2, -2, -3, -1, -5, -1, -3, -2, -2, -1,\nonumber\\& & \; \;
% -12, -1, -2, -2, -3, -1, -5, -1, -3, -2, -2, -1, -12
-12//-11//-12//-12//-12//-12//-12//-12//
 -1, -2, -2,
 -3,\nonumber\\& & \; \; -1,%\nonumber\\& & \; \;
 -5, -1, -3, -1) \nonumber
\end{eqnarray}

\section{Bases giving $U(1)$ factors}
\label{sec:appendix-abelian}

This appendix contains a list of the 13 bases over which a generic
elliptic fibration has Mordell-Weil group of nonzero rank $r$.  
Note that two of these (numbers  4, 5) also appear in the list
of Calabi-Yau's with new Hodge numbers.

\begin{eqnarray}
r = 1: &  & N = 3, 
n_0 = -2, n_\infty = -3; \; \;T = 15;   \; \;
\ho = 34,\htt = 10
\\
\small
{\rm chain}\ 1: &  &  (-2, -1, -2) \nonumber\\
{\rm chain}\ 2: &  & (-2, -2, -1, -4, -1) \nonumber\\
{\rm chain}\ 3: &  & (-2, -2, -2, -2, -2 -1, -8, -1, -2)  \nonumber
\normalsize
\end{eqnarray}

\begin{eqnarray}
r = 2: &  & N = 3, 
n_0 = -1, n_\infty = -2; \; \;T = 12;   \; \;
\ho = 24,\htt = 12
\\
\small
{\rm chain}\ 1: &  &  (-2, -1, -2) \nonumber\\
{\rm chain}\ 2: &  & (-3, -1, -2, -2) \nonumber\\
{\rm chain}\ 3: &  & (-6, -1, -2, -2, -2, -2, -2)  \nonumber
\normalsize
\end{eqnarray}

\begin{eqnarray}
r = 2: &  & N = 3, 
n_0 = -2, n_\infty = -6; \; \;T = 18;   \; \;
\ho = 46,\htt = 10
\\
\small
{\rm chain}\ 1: &  &  (-2, -1, -3, -1) \nonumber\\
{\rm chain}\ 2: &  & (-2, -2, -2, -1, -6, -1, -3, -1) \nonumber\\
{\rm chain}\ 3: &  & (-2, -2, -2, -1, -6, -1, -3, -1)\nonumber
\normalsize
\end{eqnarray}

\begin{eqnarray}
r = 2: &  & N = 3, 
n_0 = -5, n_\infty = -6; \; \;T = 21;   \; \;
\ho = 61,\htt = 1
\\
\small
{\rm chain}\ 1: &  &  (-1, -3, -1, -3, -1) \nonumber\\
{\rm chain}\ 2: &  & (-1, -3, -2, -2, -1, -6, -1, -3, -1) \nonumber\\
{\rm chain}\ 3: &  & (-1, -3, -2, -2, -1, -6, -1, -3, -1)\nonumber
\normalsize
\end{eqnarray}

\begin{eqnarray}
r = 3: &  & N = 3, 
n_0 = -4, n_\infty = -8; \; \;T = 19;   \; \;
\ho = 62,\htt = 2
\\
\small
{\rm chain}\ 1: &  &  (-1, -4, -1, -2, -3, -2, -1) \nonumber\\
{\rm chain}\ 2: &  &  (-1, -4, -1, -2, -3, -2, -1) \nonumber\\
{\rm chain}\ 3: &  &  (-1, -4, -1, -2, -3, -2, -1) \nonumber
\normalsize
\end{eqnarray}

\begin{eqnarray}
r = 4: &  & N = 3, 
n_0 = -1, n_\infty = -2; \; \;T = 11;   \; \;
\ho =  25,\htt = 13
\\
\small
{\rm chain}\ 1: &  &  (-2, -1, -2) \nonumber\\
{\rm chain}\ 2: &  & (-4, -1, -2, -2, -2) \nonumber\\
{\rm chain}\ 3: &  & (-4, -1, -2, -2, -2)  \nonumber
\normalsize
\end{eqnarray}

\begin{eqnarray}
r = 4: &  & N = 3, 
n_0 = -1, n_\infty = -5 \; \;T = 14;   \; \;
\ho = 40,\htt = 4
\\
\small
{\rm chain}\ 1: &  &  (-2, -1, -3, -1) \nonumber\\
{\rm chain}\ 2: &  &  (-4, -1, -2, -2, -3, -1) \nonumber\\
{\rm chain}\ 3: &  &  (-4, -1, -2, -2, -3, -1)  \nonumber
\normalsize
\end{eqnarray}

\begin{eqnarray}
r = 4: &  & N = 4, 
n_0 = -6, n_\infty = -6; \; \;T = 17;   \; \;
\ho = 51,\htt = 3
\\
\small
{\rm chain}\ 1: &  &  (-1, -3, -1, -3, -1) \nonumber\\
{\rm chain}\ 2: &  & (-1, -3, -1, -3, -1) \nonumber\\
{\rm chain}\ 3: &  &  (-1, -3, -1, -3, -1) \nonumber\\
{\rm chain}\ 4: &  & (-1, -3, -1, -3, -1)  \nonumber
\normalsize
\end{eqnarray}

\begin{eqnarray}
r = 4: &  & N = 3, 
n_0 = -2, n_\infty = -4; \; \;T = 13;   \; \;
\ho = 35,\htt =  11
\\
\small
{\rm chain}\ 1: &  &  (-2, -2, -1, -4, -1) \nonumber\\
{\rm chain}\ 2: &  & (-2, -2, -1, -4, -1) \nonumber\\
{\rm chain}\ 3: &  & (-2, -2, -1, -4, -1)\nonumber
\normalsize
\end{eqnarray}

\begin{eqnarray}
r = 5: &  & N = 3, 
n_0 = -1, n_\infty = -8; \; \;T = 16;   \; \;
\ho = 51,\htt = 3
\\
\small
{\rm chain}\ 1: &  &  (-3, -1, -2, -3, -2, -1) \nonumber\\
{\rm chain}\ 2: &  &  (-3, -1, -2, -3, -2, -1) \nonumber\\
{\rm chain}\ 3: &  &  (-3, -1, -2, -3, -2, -1)  \nonumber
\normalsize
\end{eqnarray}

\begin{eqnarray}
r = 6: &  & N = 3, 
n_0 = -1, n_\infty = -2; \; \;T = 10;   \; \;
\ho = 24,\htt = 12
\\
\small
{\rm chain}\ 1: &  &  (-3, -1, -2, -2) \nonumber\\
{\rm chain}\ 2: &  & (-3, -1, -2, -2) \nonumber\\
{\rm chain}\ 3: &  & (-3, -1, -2, -2)  \nonumber
\normalsize
\end{eqnarray}

\begin{eqnarray}
r = 6: &  & N = 4, 
n_0 = -2, n_\infty = -6; \; \;T = 13;   \; \;
\ho = 35,\htt = 11
\\
\small
{\rm chain}\ 1: &  &  (-2, -1, -3, -1) \nonumber\\
{\rm chain}\ 2: &  & (-2, -1, -3, -1) \nonumber\\
{\rm chain}\ 3: &  &  (-2, -1, -3, -1) \nonumber\\
{\rm chain}\ 4: &  & (-2, -1, -3, -1)  \nonumber
\normalsize
\end{eqnarray}

\begin{eqnarray}
r = 8: &  & N = 4, 
n_0 = -2, n_\infty = -2; \; \;T = 9;   \; \;
\ho = 19,\htt = 19
\\
\small
{\rm chain}\ 1: &  &  (-2, -1, -2) \nonumber\\
{\rm chain}\ 2: &  & (-2, -1, -2) \nonumber\\
{\rm chain}\ 3: &  &  (-2, -1, -2) \nonumber\\
{\rm chain}\ 4: &  & (-2, -1, -2)  \nonumber
\normalsize
\end{eqnarray}

Note that this last example is a special limit of $dP_9$ with two $-2$
clusters of type 
$I_0^* (\hat{D}_4)$.

\end{document}